\documentclass[twocolumn,trackchanges]{aastex631}

\usepackage{amsmath}
\usepackage{multirow}
\usepackage{makecell}
\usepackage{graphicx}

\begin{document}

\title{Identification of Star Clusters in M31 from PAndAS Images Based on Deep Learning}

\correspondingauthor{Bingqiu Chen}
\email{bchen@ynu.edu.cn}

\author[0009-0003-2888-6317]{Baisong Zhang}
\affiliation{South-Western Institute for Astronomy Research, Yunnan University, Kunming, Yunnan 650091, China}

\author[0000-0003-2472-4903]{Bingqiu Chen}
\affiliation{South-Western Institute for Astronomy Research, Yunnan University, Kunming, Yunnan 650091, China}

\author[0000-0003-2471-2363]{Haibo Yuan}
\affiliation{School of Physics and Astronomy, Beijing Normal University, Beijing 100875, China}
\affiliation{Institute for Frontiers in Astronomy and Astrophysics, Beijing Normal University, Beijing 102206, China}

\author[0009-0007-5623-2475]{Pinjian Chen}
\affiliation{Department of Astronomy, Yunnan University, Kunming, Yunnan 650500, China}
\affiliation{CAS Key Laboratory of Optical Astronomy, National Astronomical Observatories, Chinese Academy of Sciences, Beijing 100101, China}
\affiliation{School of Astronomy and Space Science, University of Chinese Academy of Sciences, Beijing 100049, China}

\author[0000-0002-5909-6233]{Shoucheng Wang}
\affiliation{CAS Key Laboratory of Theoretical Physics, Institute of Theoretical Physics, Chinese Academy of Sciences, Beijing 100190, China}

\author[0000-0003-3618-9960]{Lunwei Zhang}
\affiliation{South-Western Institute for Astronomy Research, Yunnan University, Kunming, Yunnan 650091, China}

\author[0000-0003-1218-8699]{Yi Ren}
\affiliation{Department of Astronomy, College of Physics and Electronic Engineering, Qilu Normal University, Jinan, Shandong 250200, China}

\author[0000-0001-5737-6445]{Helong Guo}
\affiliation{South-Western Institute for Astronomy Research, Yunnan University, Kunming, Yunnan 650091, China}

\begin{abstract}
The identification of star clusters holds significant importance in studying galaxy formation and evolution history. However, the task of swiftly and accurately identifying star clusters from vast amounts of photometric images presents an immense challenge. To address these difficulties, we employ deep learning models for image classification to identify young disk star clusters in M31 from the Pan-Andromeda Archaeological Survey (PAndAS) images. For training, validation, and testing, we utilize the Panchromatic Hubble Andromeda Treasury (PHAT) survey catalogs. We evaluate the performance of various deep learning models, using different classification thresholds and limiting magnitudes. Our findings indicate that the ResNet-50 model exhibits the highest overall accuracy. 
Moreover, using brighter limiting magnitudes and increasing the classification thresholds can effectively enhance the accuracy and precision of cluster identification. Through our experiments, we found that the model achieves optimal performance when the limiting magnitude is set to brighter than 21\,mag. Based on this, we constructed a training dataset with magnitudes less than 21\,mag and trained a second ResNet-50 model. This model achieved a purity of 89.30\%, a recall of 73.55\%, and an F1 score of 80.66\% when the classification threshold was set to 0.669. Applying the second model to all sources in the PAndAS fields within a projected radius of 30\,kpc from the center of M31, we identified 2,228 new unique star cluster candidates. We conducted visual inspections to validate the results produced by our automated methods, and we ultimately obtained 1,057 star cluster candidates, of which 745 are newly identified.
\end{abstract}

\section{Introduction} \label{sec:intro}

Star clusters hold significant importance in tracing the historical progression of galaxy formation and evolution, making them invaluable tools in understanding the assembly and evolution history of galaxies. In particular, studying star clusters in the Andromeda galaxy M31, the nearest large spiral galaxy, is of great importance. The identification of M31 star clusters is fundamental to these studies. The continuous advancement of astronomical telescopes has provided us with a wealth of photometric data. However, efficiently and accurately identifying star clusters from such vast amounts of photometric data remains an extremely challenging task.

Numerous studies have been conducted to identify star clusters in M31, and we present a brief overview of some recent works. \cite{hodge2010photometric} examined Hubble Space Telescope (HST) WFPC2 images to discover 77 new star clusters in active star-formation regions of M31. \cite{johnson2012phat} conducted a visual search of high spatial resolution HST images from the Panchromatic Hubble Andromeda Treasury (PHAT) survey, resulting in the identification of 601 clusters. In a subsequent study, \cite{johnson2015phat} utilized visual image classification performed by the Andromeda Project citizen science website to identify 2,753 clusters and 2,270 background galaxies. \cite{di2014search} visually examined Sloan Digital Sky Survey (SDSS) images and identified seven globular cluster candidates in the M31 halo. \cite{huxor2014outer} discovered 59 globular clusters and two candidates in the halo of M31 through visual inspection of MegaCam images obtained from the Pan-Andromeda Archaeological Survey (PAndAS) conducted with the Canada-France-Hawaii Telescope (CFHT). \cite{chen2015lamost} employed analysis of LAMOST spectra and morphological information from SDSS images to identify 28 globular cluster candidates in M31.

Most previous studies relied heavily on visual inspection, which demanded substantial time and effort. With the advent of modern wide-field galaxy surveys, tens of millions of source images are now available, rendering traditional visual inspection methods increasingly inefficient. In recent years, artificial intelligence (AI) techniques have emerged as a powerful tool for the automatic classification of star cluster candidates. By leveraging AI, it is possible to perform an initial screening of candidate star clusters, greatly reducing the workload associated with manual visual inspections while significantly improving both efficiency and consistency. 
\cite{bialopetravivcius2020study} utilized a convolutional neural network (CNN) to train simulated star clusters and successfully detected 3,380 star cluster candidates in M83 from HST observations. 
\cite{wei2020deep} employed neural network models and deep transfer learning techniques for morphological classification of compact star clusters in nearby galaxies in the homogeneous data sets of human-labelled star cluster images from the HST, and proved the good performance of these models. 
\cite{perez2021starcnet} proposed StarCNet, a multi-scale CNN method and applied it to multicolor images obtained from HST observations to identify star cluster candidates in nearby galaxies. \cite{wang2022identification} proposed a method utilizing CNN to identify new star clusters in M31 using PAndAS images. After a visual check, they identified 117 new candidates from approximately 5000 CNN candidates.
\cite{hannon2023star} used HST ultraviolet (UV)-optical imaging of over 20,000 sources in 23 galaxies from the PHANGS survey to show the performance of the deep transfer learning techniques for star cluster morphological classification, they found distance-dependent models and distance-independent models had little impact on the classification results. \cite{wang2023searching} trained two random forest classifiers using the catalogs of Gaia Early Data Release 3 and Pan-STARRS 1. They then conducted visual inspection using PAndAS images to eliminate non-cluster sources, resulting in 50 globular cluster candidates from $\sim$ 2000 model predicted candidates.

Many of the aforementioned studies rely on shallow CNNs, which still produce a significant number of false positives, resulting in a heavy burden for subsequent visual inspection tasks. Our goal is to obtain a star cluster sample with higher purity, thereby paving the way for fully automated star cluster classification. In recent years, deep learning-based image classification algorithms have matured and demonstrated significant advantages, particularly in industrial applications, offering promising solutions for improving classification accuracy and efficiency. Early image classification methods heavily relied on handcrafted feature extraction algorithms such as Histogram of Oriented Gradients (HOG) features \citep{dalal2005histograms}, Scale-Invariant Feature Transform (SIFT) features \citep{lowe1999object}, and Harr-like features~\citep{viola2004robust}. These traditional approaches required manual selection and design of task-specific features, followed by the use of machine learning algorithms like Support Vector Machines (SVM; \citealp{noble2006support}) for classification. The advancements in Graphics Processing Units (GPUs) and the availability of large-scale datasets have led to breakthroughs in image classification using deep learning. CNNs have enabled models to automatically learn representations from raw pixels to high-level features. The emergence of AlexNet ~\citep{krizhevsky2012imagenet} marked the beginning of the success of deep learning in image classification. AlexNet utilized multiple convolutional layers, fully connected layers, Rectified Linear Unit (ReLU) activation, and Dropout to address overfitting. VGG~\citep{simonyan2014very} introduced a very deep network structure with consecutive convolutions using small-sized filters, enabling the capture of more detailed feature representations. GoogLeNet~\citep{szegedy2015going} introduced the Inception module, which captured features at different abstraction levels by employing parallel filters of various sizes. The stacking of these modules allowed networks to achieve both efficiency and performance. ResNet~\citep{he2016deep} introduced the concept of residual learning by incorporating skip connections. These connections directly add the input to the output of intermediate layers, thereby addressing the issues of gradient vanishing and degradation during the training of deep networks. DenseNet~\citep{huang2017densely} improved upon the architecture of ResNet by introducing dense connections. In DenseNet, each layer is connected to all preceding layers, facilitating better information propagation and reuse, thereby enhancing feature utilization efficiency.

In this study, our objective is to employ deep learning methods for the automatic classification of disk young clusters in M31 using PAndAS images. Disk young clusters pose a greater challenge for identification compared to globular clusters due to their fainter nature, making them harder to detect in ground observations. To train and validate our deep learning models, we utilize the catalogs from the HST. 
We evaluated the classification performance by analyzing the purity, recall, and F1 score under various conditions, including limiting magnitudes, classification thresholds, projected distance, sizes, color ($g - i$), and distances from the centers of CCDs. By setting the limiting magnitude to 21\,mag and the classification threshold to 0.669, the model achieved a classification purity exceeding 80\% for star clusters. To further enhance the purity of the catalog, we focused on data with $g$-band magnitudes less than 21\,mag and trained a second ResNet-50 model. Applying this second model, we identified 2,228 independent star cluster candidates. We conducted visual inspections to validate the results produced by our automated methods. Ultimately, 1,057 star cluster candidates were obtained through visual inspection, including 745 newly identified clusters.

\section{Data} \label{sec:Data}

\subsection{PAndAS Images}

In this study, we utilize the PAndAS $g$- and $i$-band images obtained from the CFHT to train our deep learning models and identify new star cluster candidates in M31. PAndAS is a large-scale photometric survey specifically designed to explore the structure and content of the M31 and its neighboring Triangulum galaxy M33 \citep{mcconnachie2018large}.  The survey was conducted between 2003 and 2010, using the MegaCam wide-field camera \citep{boulade2003megacam} which comprises 36 CCDs, each with 2048 $\times$ 4612 pixels. The camera has an effective field of view of 0.96 $\times$ 0.94\,deg$^2$ and a pixel scale of 0.187”/pixel. The typical seeing values for the $g$- and $i$-bands in the PAndAS survey are 0.67" and 0.60", respectively \citep{huxor2014outer}. In this work, we use the processed stacked images that were processed by the Cambridge Astronomical Survey Unit (CASU) from the PAndAS VOspace\footnote{\url{https://www.canfar.net/storage/vault/list/PANDAS/PUBLIC}}, including image processing, calibration, and photometric measurements.

Based on the PAndAS sources catalog, we extract images of the individual sources. To obtain a standardized size, we extend 28 pixels in each direction (up, down, left, and right) from the center coordinates ($x$, $y$) of the source, resulting in a 56 $\times$ 56 pixel image. However, for sources located at the edges of the images, we could not obtain complete images. To address this, we use padding by filling the exceeding parts with a pixel value of 0. 
The proportion of training and test images with incomplete pixel coverage is very small and almost negligible. We include these images for two main reasons: to expand some of the original positive samples and to avoid discarding similar sources in subsequent star cluster identifications. 
However, the resulting images are relatively small for deep learning image classification models, which may affect classification performance. The standard input size for deep classification models is 224 $\times$ 224. Therefore, we enlarged the images using interpolation to achieve a final size of 224 $\times$ 224 pixels, which matches the input size required by the deep learning models.

\begin{figure}
	\centering
	\includegraphics[width=1.0\linewidth]{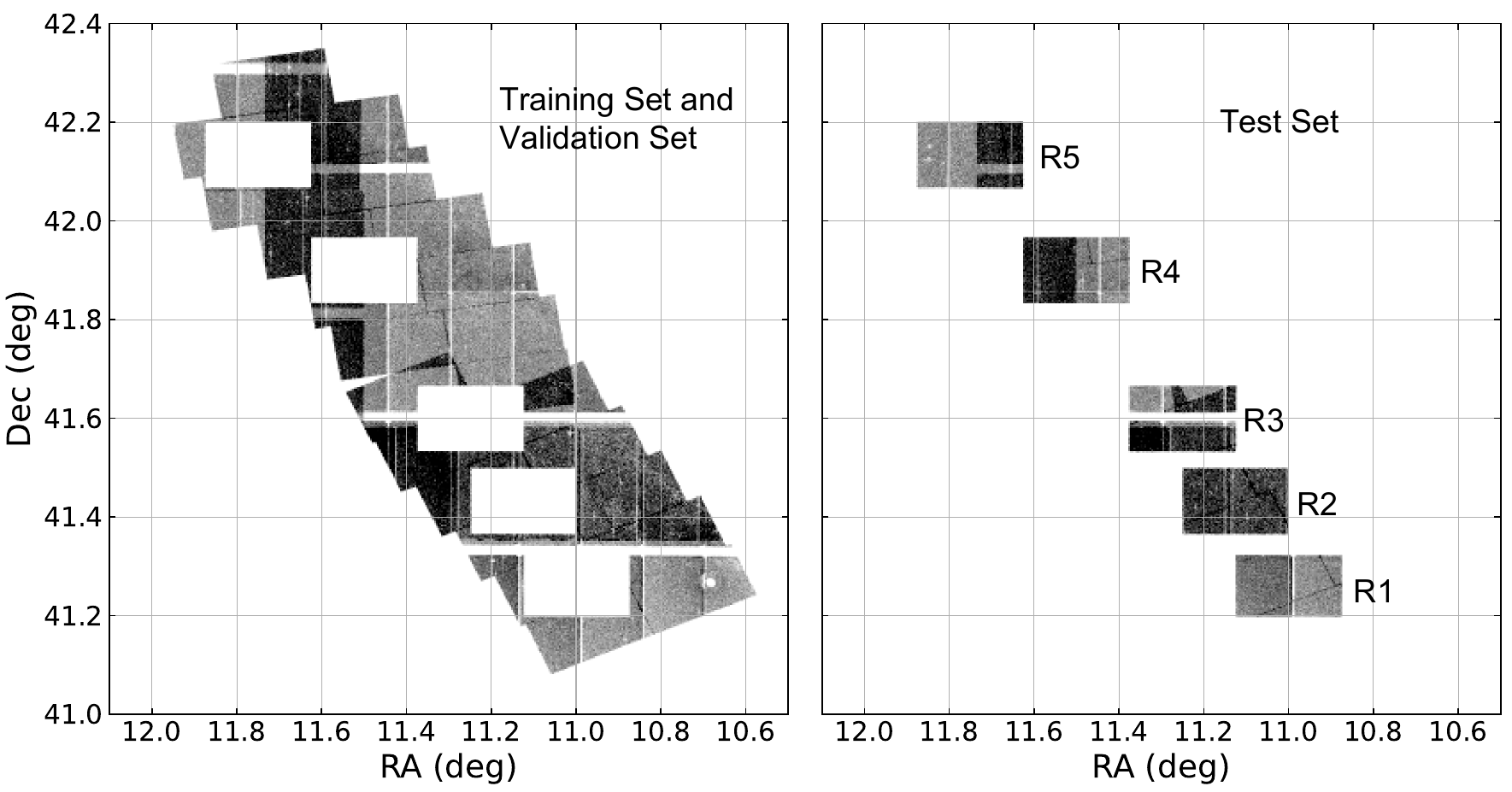}
	\caption{The regions where the training and validation sets (Region a) and the test set (Region b) are located.}
	\label{datasetRegion}
\end{figure}

Similar to \cite{liu2019classification} and \cite{wang2022identification}, we employ the standard z-score method to preprocess the PAndAS image stamps. The z-score method is defined as follows:
\begin{equation}
	\begin{split}
		g_1 = \frac{g - \overline{g}}{\sigma_g},\\
		i_1 = \frac{i - \overline{i}}{\sigma_i}.\
	\end{split}
\end{equation}
Here, $g$ and $i$ represent the flux values of individual pixels, while $\overline{g}$ and $\overline{i}$ denote the mean flux values of the image stamps, respectively. Additionally, $\sigma_g$ and $\sigma_i$ represent the standard deviations. Standardizing with z-scores is a widely adopted technique in machine learning due to its efficiency.

For training the deep learning models, we needed to synthesize color images using data from three bands/channels. As we have only two bands images, to ensure that the third channel appears white, we filled all the blue channels of the color images with a pixel value of 255, which can be defined as $b_1 = 255$. RGB images were synthesized using $g_1$, $i_1$, and $b_1$, corresponding to the primary colors of red, green, and blue, respectively. 
Similarly, the use of the RGB three channels ensures compatibility with the input format required by deep learning models.

\subsection{Positive and Negative Samples}

To train our deep learning models, we require a substantial number of positive and negative sample images. As positive samples, we utilize the confirmed star cluster catalog derived from the PHAT survey project conducted by \citet{johnson2015phat}. This catalog was constructed through the classification of 20,000 images by tens of thousands of volunteers, resulting in 1.82 million classifications and the identification of 2,753 star clusters. In this study, we make the assumption that the star cluster catalog in \citet{johnson2015phat} is complete and free from contamination. However, it should be noted that \citet{johnson2015phat} demonstrated that their catalog is 50\% complete for clusters with masses equal to 500\,$M_{\odot}$ for ages younger than 100\,Myr. Despite this, the PHAT images, obtained from a space telescope, possess significantly higher spatial resolution and quality compared to the PAndAS images. Therefore, we can consider a PHAT-like result achieved with ground-based PAndAS data as complete and uncontaminated.

For the negative samples, we incorporate the confirmed galaxy catalog from the PHAT survey project \citep{johnson2015phat} and the sources cataloged by Nelson Caldwell\footnote{\url{https://www.cfa.harvard.edu/oir/eg/m31clusters/M31_Hectospec.html}} in the PHAT region that are not star clusters. These sources encompass H~\textsc{ii} regions, planetary nebulae, stars, and symbiotic stars in M31. Furthermore, we randomly select sources within the PHAT area that are not classified as star clusters in \cite{johnson2015phat} as negative samples, and these sources are the main components of negative samples.

\begin{table*}
	\caption{Dataset statistics.}
	\label{statistics}
	\setlength{\tabcolsep}{1.2mm}{
		\renewcommand{\arraystretch}{1.2}
		\begin{tabular}{ccccccccc}
			\hline
			Class & Training set & Validation Set & Test Set & Test R1 & Test R2 & Test R3 & Test R4 & Test R5 \\
			\hline
			SC & 105,000 & 14,952 & 1,060 & 172 & 335 & 203 & 251 & 99 \\
			nonSC & 104,963 & 14,989 & 376,268 & 57,462 & 91,824 & 74,900 & 80,701 & 71,381 \\
			\hline
	\end{tabular}}
\end{table*}

\subsection{Training, Validation and Test Sets}

We randomly select five subfields from the PHAT region. In our study, we distinguish the remaining fields as "Region a" and the selected fields as "Region b". The positive and negative samples located in Region a are used as the training and validation sets, while those in Region b are designated as the test set. The validation set is independent of the test set and is solely employed for hyperparameter adjustment to prevent any bias in hyperparameter selection. Fig.~\ref{datasetRegion} displays the regions where the training and validation datasets (Region a) and the test dataset (Region b) are situated. The subfields within Region b are labeled as R1 to R5, with R1 being the subfield closest to the center of M31 and R5 being the subfield farthest away, arranged in order of increasing radius. The irregular source density distribution in Fig.~\ref{datasetRegion} is due to higher data density in overlapping PAndAS photometry regions. 

We have obtained a total of 2,142 star clusters and 1,896 galaxies from the catalogs of \cite{johnson2015phat} in Region a. Due to the small quantity of Region a, we obtained 2,060 non-clusters (including 358 H~{\sc ii} regions, 658 planetary nebulae, 1,013 stars, and 31 symbiotic stars) from the Caldwell catalog in all M31 Region. Note that the saturated or zero pixels have not been masked in this work. Since these pixels contain information resulting from the observation. 
For the star cluster catalog, we cross-matched sources with a 1$^{\prime\prime}$ radius in the PAndAS catalogs. Most sources matched well, although not all were perfect. We used the $XY$ coordinates from the PAndAS catalogs to create cutouts. For the training sample, we cross-matched the individual catalogs (e.g., star cluster and galaxy catalog from \cite{johnson2015phat}, star, PNe,  H~{\sc ii}, and symbiotic stars catalog from Caldwell) and only two common stars were found (cross-matching of the galaxy catalog with the  H~{\sc ii} catalog) across these catalogs. We used the field-specific individual object catalogs as input for the PAndAS data. The adjacent PAndAS regions overlap, and we did not eliminate duplicate data, treating overlapping data as separate observations. The number of star clusters for training and validation sets is only 2,142, which is relatively small for training a deep learning model. To address this, we have augmented the data artificially to increase the number of positive samples. Initially, we divided the 2,142 star clusters in Region a into a training set and a validation set in a ratio of 7:1. Subsequently, we employed various methods such as rotation, flipping, translation, and Gaussian enhancement to expand the dataset of star cluster images. By applying these techniques, we obtained a total of 119,952 star cluster images as positive samples for the training and validation sets.

To maintain a balance between positive and negative samples, we randomly selected sources that were not identified as star clusters in the \cite{johnson2015phat} catalogs as non-star cluster sources, and added them to the negative samples. Consequently, we achieved an equal number of 119,952 negative samples for the training and validation sets. For the test set, we refrained from performing any data augmentation, given that the comparison between clusters and non-clusters in real scenarios is inherently disparate. The test set comprised 377,328 images, including 1,060 star clusters and 376,238 non-clusters. The negative samples in the test set were not randomly selected, they included all remaining data after selecting the best cross-matched star clusters as positive samples. We first selected five regions, cross-matched them with the \cite{johnson2015phat} star cluster catalog, and designated the best matches as positive samples, with the rest as negative samples. 
Table~\ref{statistics} presents the dataset statistics, where the SC class denotes star clusters, and the nonSC class represents other sources excluding star clusters.

\section{Residual Networks} \label{sec:method}

In the rapidly advancing field of deep learning, significant progress has been made in image classification, resulting in higher accuracy and improved generalization. Existing image classification models have demonstrated excellent performance even when the distributions of training and test sets are independent to each others \citep{krizhevsky2012imagenet, simonyan2014very, szegedy2015going}. However, increasing the depth of deep learning networks does not necessarily lead to improved classification performance. Instead, it can result in slower network convergence and decreased accuracy. Even expanding the dataset to address overfitting does not improve classification performance and accuracy. As the number of layers in deep neural networks increases, the expressive power of the model tends to saturate or decrease due to the problems of gradient vanishing and degradation caused by network deepening.

To alleviate the issues of gradient vanishing, Residual Network (ResNet; \citealp{he2016deep}) has emerged as a highly effective approach. ResNet enables the training of deep networks by implementing residual connections, where the input of a unit is directly added to its output before activation. This approach mitigates the problems of gradient vanishing and degradation, allowing deep neural networks to achieve greater improvements in expressive power.

\begin{figure}
	\centering
	\includegraphics[width=1.0\linewidth]{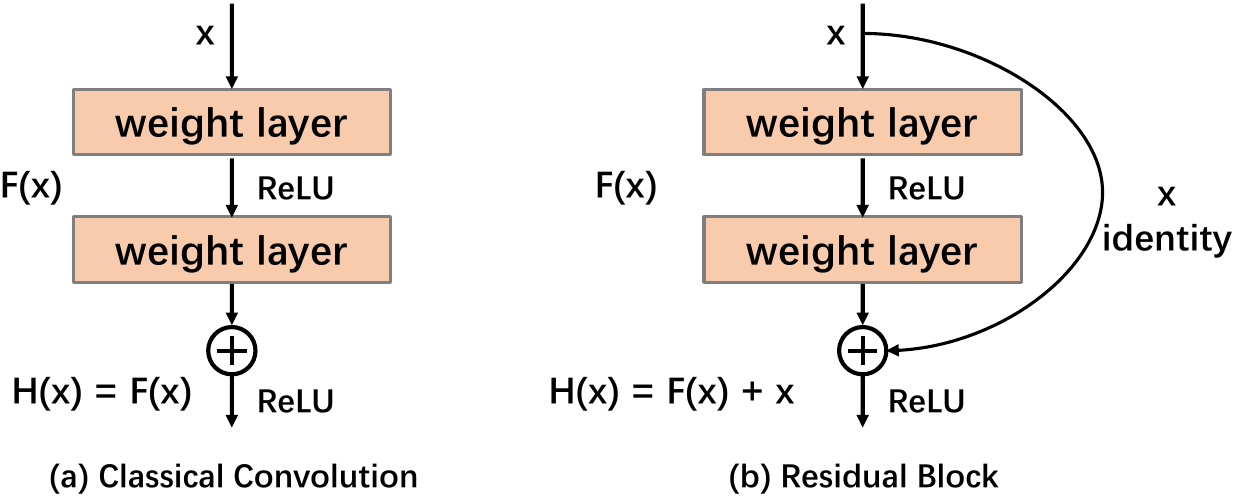}
	\caption{Sketch Map of classical convolutional network and residual network.}
	\label{ccrb}
\end{figure}

Fig.~\ref{ccrb} illustrates the concept of Residual Network. The input feature map is denoted as $x$, the desired underlying mapping as $H(x)$, and the current map held in the parameters as $F(x)$. In classical convolution layers, $H(x) = F(x)$. However, in ResNet, the goal is to learn a residual mapping $H(x)$ such that $F(x) = H(x) - x$, which recasts the desired mapping into $F(x) + x$, where $x$ represents the identity mapping. Shortcut connections are used to skip several convolution layers, enabling the formulation of $H(x) = F(x) + x$. Optimizing the residual mapping is simpler than optimizing the original, unreferenced mapping.

If newly added layers perform poorly, they can be bypassed through residual connections by setting the weight parameters of those layers to 0. This ensures that excellent layers are retained while ineffective parts are skipped. Therefore, regardless of the number of layers in the network, the overall performance of the model does not decrease. Increasing the number of layers can enhance the performance of the model, making this an important feature.

\begin{figure*}
	\centering
	\includegraphics[width=0.6\linewidth]{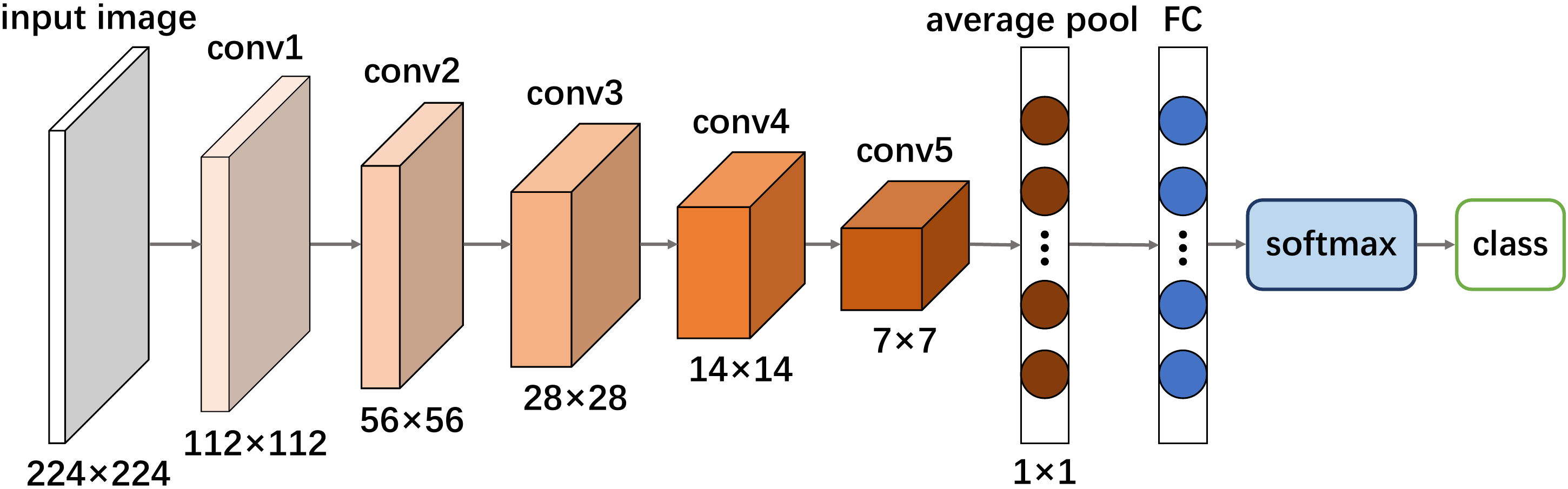}
	\caption{The structure of ResNet-50.}
	\label{res50f}
\end{figure*}

\begin{table*}
	\centering
	%	\raggedright
	\caption{Architectural details of ResNet-50.}
	\label{res50t}
	\setlength{\tabcolsep}{3.8mm}{
		\renewcommand{\arraystretch}{1.2}
		%	\scalebox{1}{}
		\begin{tabular}{cccc}
			\hline
			&Input Size&Output Size&Kernel\\
			\hline
			conv1&$224\times224\times3$&$112\times112\times64$&$7\times7$, 64, stride 2\\
			\hline
			\multirow{2}*{conv2}&$112\times112\times64$&$56\times56\times64$&$3\times3$ max pool, stride 2\\
			\cline{2-4}
			%				\midrule{2-4}[0.5pt]
			&$56\times56\times64$&$56\times56\times256$& $\left[ \makecell{1\times1, 64 \\ 3\times3, 64 \\ 1\times1, 256} \right]\times3$\\
			\hline
			conv3&$56\times56\times256$&$28\times28\times512$& $\left[ \makecell{1\times1, 128 \\ 3\times3, 128 \\ 1\times1, 512} \right]\times4$\\
			\hline
			conv4&$28\times28\times512$&$14\times14\times1024$& $\left[ \makecell{1\times1, 256 \\ 3\times3, 256 \\ 1\times1, 1024} \right]\times6$\\
			\hline
			conv5&$14\times14\times1024$&$7\times7\times2048$& $\left[ \makecell{1\times1, 512 \\ 3\times3, 512 \\ 1\times1, 2048} \right]\times3$\\
			\hline
			average pool&$7\times7\times2048$&$1\times1\times2048$&-\\
			\hline
			fc1000&$1\times1\times2048$&$1\times1\times1000$&-\\
			\hline
			\multicolumn{4}{c}{softmax}\\
			\hline
	\end{tabular}}
\end{table*}

\subsection{The ResNet-50 Structure}

In our study, we have conducted experiments using three different ResNet models: ResNet-18, ResNet-50 and ResNet-101\citep{he2016deep}, and one deep CNN model: VGG-16 \citep{simonyan2014very}. As demonstrated in Sect.~4.3, the ResNet-50 model outperforms the others overall, so we choose to adopt it as the model for our current work. Here, we provide a brief introduction to the structure of ResNet-50. The architecture of ResNet-50 is depicted in Fig.~\ref{res50f}, which includes input images, five convolutional blocks, one average pooling layer, one fully connected layer, and the softmax function. More detailed information on the parameters of ResNet-50 is listed in Table~\ref{res50t}.

The input image undergoes feature extraction using a 7 $\times$ 7 convolutional kernel with a stride of 2, resulting in a halving of the width and height of the image. The extracted feature map then passes through a MaxPool layer to further reduce the image resolution. ResNet-50 is divided into four convolutional blocks, each containing multiple residual blocks. Each residual block consists of several convolutional kernels, including 1 $\times$ 1, 3 $\times$ 3, and another 1 $\times$ 1 kernels. After traversing through the four convolutional blocks, we obtain a feature map with dimensions 7 $\times$ 7 $\times$ 2048, representing the width, height, and number of channels. Subsequently, the feature map undergoes average pooling, resulting in a 1 $\times$ 1 $\times$ 2048 feature map. Following that, a fully connected layer is applied to transform the feature map into a 1 $\times$ 1 $\times$ 1000 feature map. %Finally, the softmax function is employed to calculate the classification confidence and produce the final classification results. 
Finally, The ResNet-50 passes the logits through the softmax function, converting them into a probability distribution. This distribution represents the model's classification confidence for each class, with the highest probability indicating the most likely class for the input image.

\begin{table*}[ht]
	\centering
	\caption{Comparison of performance metrics for the four deep learning models.}
	\label{4result}
	\setlength{\tabcolsep}{2.8mm}{
		\renewcommand{\arraystretch}{1.2}
		\begin{tabular}{ccccc|cccc}
			\hline
			\multirow{2}*{Models}&&\multicolumn{2}{c}{Validation Set}&&&\multicolumn{2}{c}{Test Set}&\\
			&Accuracy&Recall&Precision&F1 score&Accuracy&Recall&Precision&F1 score\\
			\hline
			VGG-16&99.30\%&98.98\%&99.62\%&99.30\%&99.74\%&59.25\%&53.63\%&56.30\%\\
			ResNet-18&99.29\%&98.86\%&\textbf{99.70\%}&99.28\%&99.72\%&51.98\%&49.95\%&50.95\%\\
			ResNet-50&99.28\%&98.86\%&\textbf{99.70\%}&99.28\%&\textbf{99.75\%}&64.43\%&\textbf{54.12\%}&\textbf{58.83\%}\\
			ResNet-101&\textbf{99.44\%}&\textbf{99.20\%}&99.68\%&\textbf{99.44\%}&99.70\%&\textbf{67.64\%}&48.12\%&56.24\%\\
			\hline
	\end{tabular}}
\end{table*}

\section{Result}

\subsection{Implementation Details and Evaluation}

The image classification network models utilized in this paper are sourced from the publicly available \texttt{mmclassification} toolbox\footnote{\url{https://github.com/open-mmlab/mmpretrain}}. All models undergo training on the training set, validation on the validation set, and testing on the test set to showcase the test results. These models are trained with an initial learning rate of $1.0\times10^{-2}$. Unless otherwise specified, the remaining hyperparameters adhere to the settings in the \texttt{mmclassification} code base.

We employ four indicators: accuracy, recall, precision (purity), and F1 score. Their respective calculation formulas are as follows:
\begin{equation}
	\text{Accuracy} = \frac{TP+TN}{TP+TN+FP+FN},
\end{equation}
\begin{equation}
	\text{Precision} = \text{Purity} = \frac{TP}{TP+FN},
\end{equation}
\begin{equation}
	\text{Recall} = \frac{TP}{TP+FP},
\end{equation}
\begin{equation}
	\text{F1\ score} = \frac{2 \times \text{Precision} \times \text{Recall}}{\text{Precision} + \text{Recall}}.
\end{equation}
Among these indicators, $TP$ (True Positive) represents a cluster classified by the model as a star cluster (i.e., a true cluster correctly classified as a star cluster by the model). $TN$ (True Negative) refers to a non-cluster classified by the model as a non-cluster (i.e., a non-cluster correctly classified as a non-cluster by the model). $FP$ (False Positive) signifies a non-cluster classified by the model as a star cluster (i.e., a non-cluster misclassified as a star cluster by the model). $FN$ (False Negative) denotes a cluster classified by the model as a non-cluster (i.e., a true cluster misclassified as a non-cluster by the model).

Accuracy reflects the ratio of correctly predicted samples to the total number of samples, Recall reflects the ratio of correctly predicted positive samples to the actual total number of positive samples, Precision reflects the ratio of correctly predicted positive samples to the total number of predicted positive samples, and the F1 score is a widely employed metric for evaluating the performance of a classification model as it combines both precision and recall. 
Morphological properties such as the half-light radius are essential for traditional cluster identifications. In our work, we use the cluster images directly for classification. As such, information like the half-light radius and other morphological properties are inherently included in the input images. These metrics do not need to be input separately and do not affect our classification results independently. 

\begin{figure}
	\centering
	\includegraphics[width=0.82\linewidth]{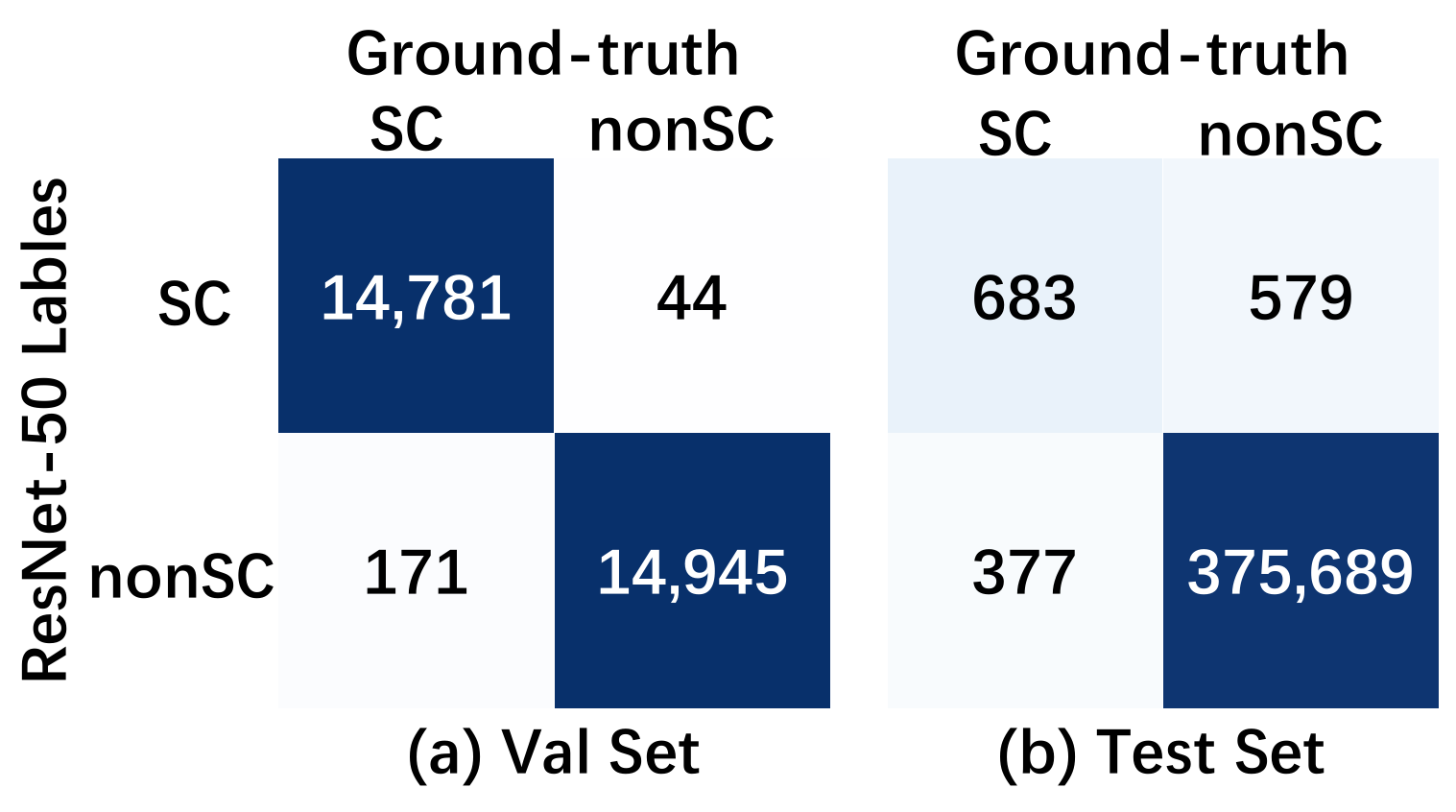}
	\caption{Confusion matrices for the ResNet-50 model on the validation set (a) and test set (b).}
	\label{res50con}
\end{figure}

\begin{figure*}
	\centering
	\includegraphics[width=0.75\linewidth]{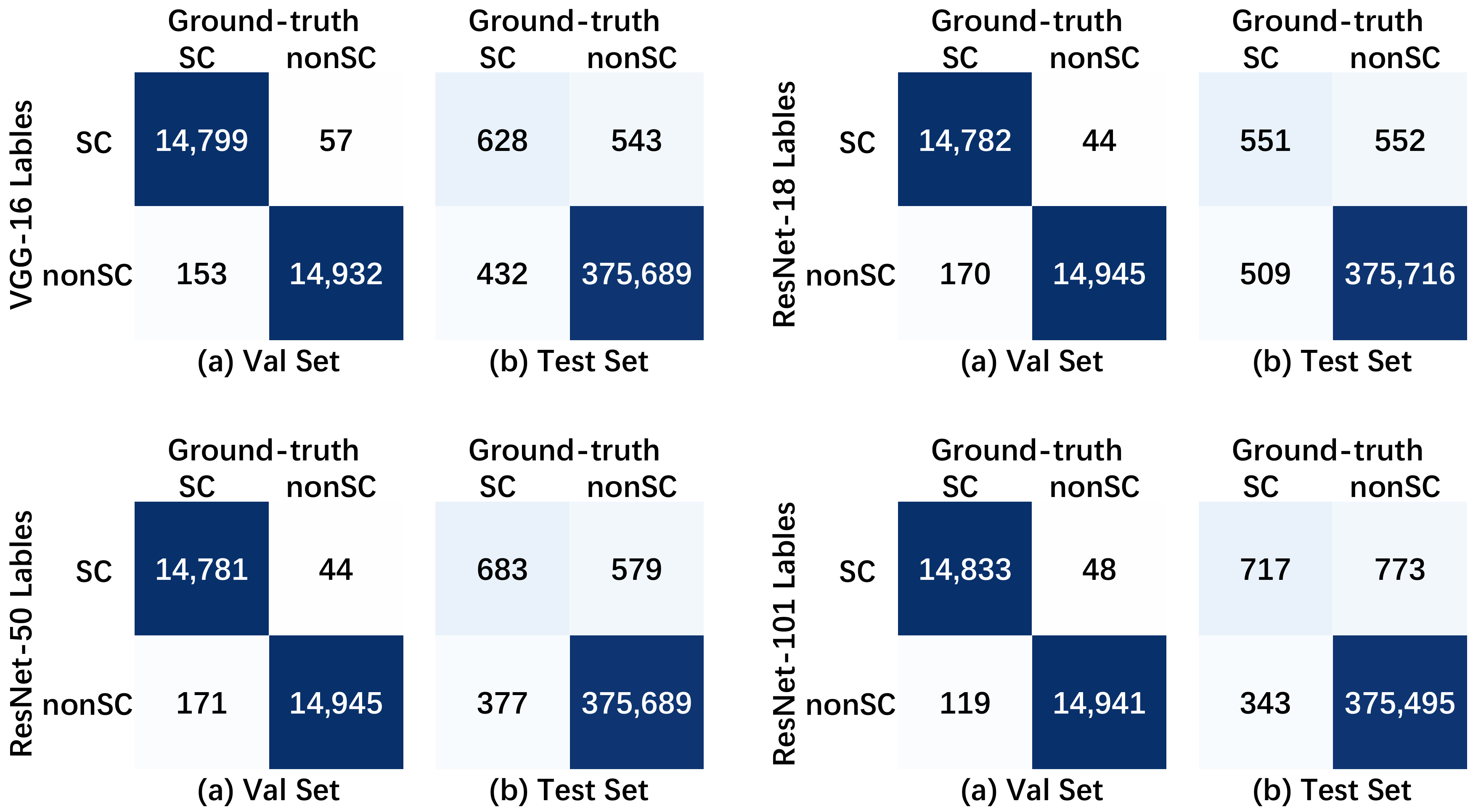}
	\caption{Confusion Matrices of the four adopted deep learning models on the validation and test sets.}
	\label{4modelcon}
\end{figure*}

\subsection{Experimental Results}

We present the confusion matrices of the ResNet-50 model on both the validation and test sets in Fig.~\ref{res50con}. By analyzing the confusion matrices, we can determine the accuracy, recall, precision, and F1 score values of the ResNet-50 model. The model exhibits excellent performance on the validation set, achieving high accuracy (99.28\%), recall (98.86\%), precision (99.70\%), and F1 score (99.28\%). 

In the case of the test set, the values for recall (64.43\%), precision (54.12\%), and F1 score (58.83\%) are relatively low. This is primarily attributed to the extreme imbalance between positive and negative samples in the test set. The majority of the sources belong to the non-cluster category, and even if a very small fraction of non-cluster sources are misclassified as clusters, the number of these misclassifications is not negligible compared to the true number of clusters. Consequently, this leads to lower values for recall, precision, and F1 score. Regarding the test set, the model misclassifies only a small fraction of instances (579) of non-cluster as star clusters, accounting for merely 0.15\% of the non-cluster category.

\subsection{Comparison between Different Models}

We present the confusion matrices for the four deep learning models adopted in this study, as shown in Fig.~\ref{4modelcon}. Additionally, Table~\ref{4result} provides the calculated accuracy, recall, precision, and F1 score values for each of these models. The results demonstrate that all four models exhibit excellent classification performance on the validation set. 

On the test set, the VGG-16, ResNet-18, ResNet-50, and ResNet-101 models correctly classify 628, 551, 683, and 717 star clusters, respectively.  While 432, 509, 377, and 343 star clusters are misclassified as non-clusters for each respective model. Among these models, the ResNet-101 model achieves the highest number of correct classifications for star clusters, while the ResNet-18 model has the fewest. Furthermore, the four models misclassify non-cluster as star clusters, with quantities of 543, 552, 579, and 773, respectively. The ResNet-101 model exhibits the highest number of misclassifications, whereas the other three models have similar quantities of misclassifications.  Overall, the confusion matrix indicates that the classification performance of the four models is comparable. 

The recall values for the VGG-16, ResNet-18, ResNet-50, and ResNet-101 models are 59.25\%, 51.98\%, 64.43\%, and 67.64\%, respectively. The ResNet-101 model exhibits the highest recall, while the ResNet-18 model has the lowest. In terms of precision, the four models achieve values of 53.63\%, 49.95\%, 54.12\%, and 58.83\%, respectively. The ResNet-50 model demonstrates the highest precision. The F1 scores for the four models are 56.30\%, 50.95\%, 58.83\%, and 56.24\%, respectively. The ResNet-50 model achieves the highest F1 score, indicating the best overall performance. The F1 scores for the VGG-16 and ResNet-101 models are nearly identical, while the ResNet-18 model exhibits a slightly lower F1 score. 

\subsection{Magnitude and Classification Threshold}\label{sec:mag_thr}

Based on the experimental results, it is observed that all models have achieved high classification accuracy for the artificially balanced validation set. However, when tested with real unbalanced data, the precision is relatively low. The test set exhibits a significant imbalance, with a ratio of 1:355 between star clusters and non-clusters. This severe imbalance has a substantial impact on the precision of the star cluster identification works. To obtain a high-purity sample of cluster candidates, it is necessary to further analyze the properties of the model-predicted cluster candidates, such as magnitude and model classification thresholds, etc. In the subsequent analyses, we will exclusively discuss the results obtained from the ResNet-50 model, unless otherwise specified.

\begin{figure}
	\centering
	\includegraphics[width=1.0\linewidth]{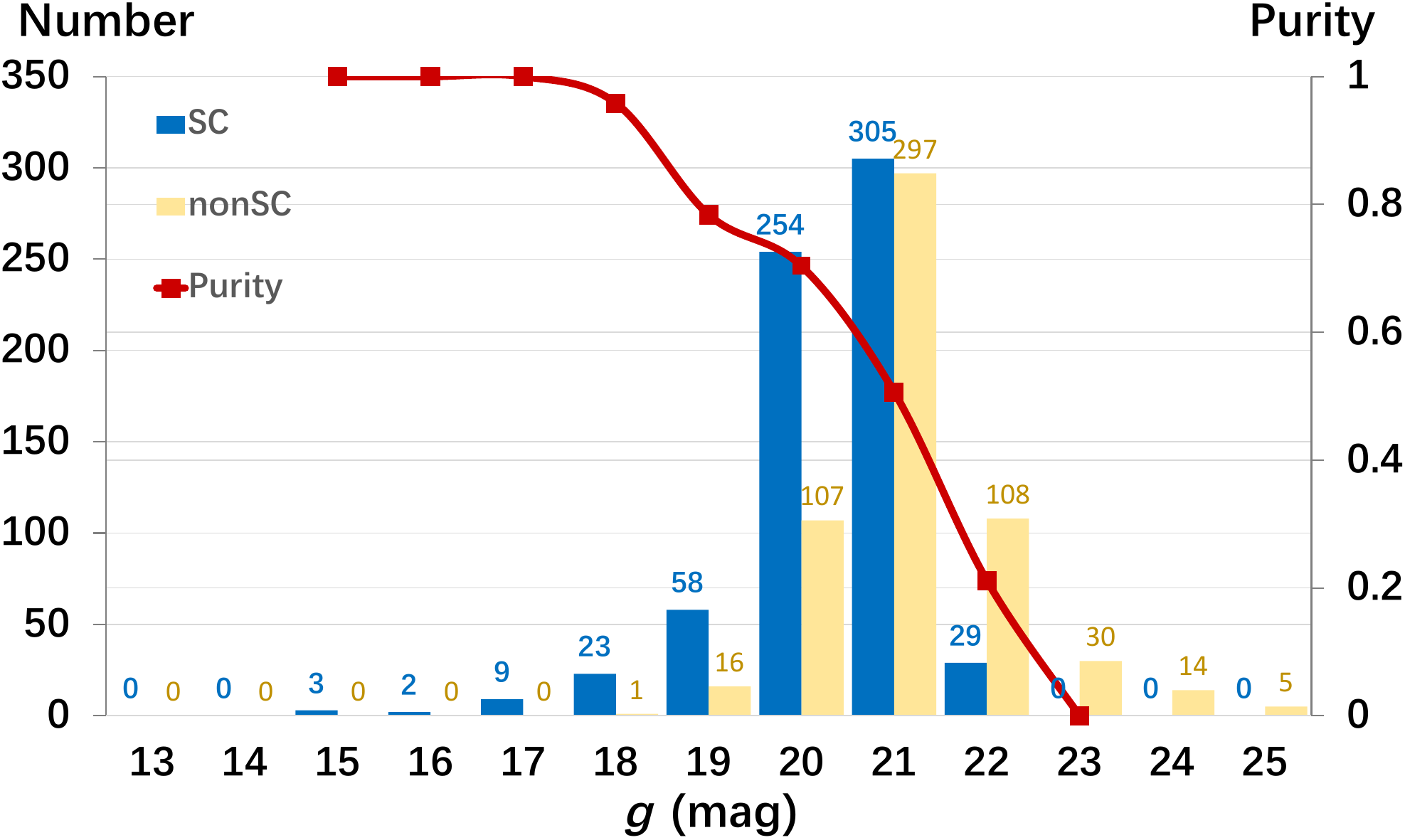}
	\caption{Variations in classification purity (red curve) and the number of correctly classified clusters as star clusters (blue bars) and misclassified non-clusters as star clusters (yellow bars) across different $g$-band magnitude bins. These measurements are obtained by evaluating the RestNet-50 model on the test set.}
	\label{mag1}
\end{figure}

We begin by examining the behaviors of the trained ResNet-50 model on the test set for sources with different magnitudes. Fig.~\ref{mag1} illustrates the variation in classification purity for star clusters based on their $g$-band magnitudes. Additionally, the Figure displays the number of model predicted clusters, including the correctly classified true clusters and the misclassified non-clusters, across various magnitude bins.

From Fig.~\ref{mag1}, it is apparent that classification purity decreases as the magnitude becomes fainter. This decline can be attributed to the lower signal-to-noise ratios associated with objects of fainter magnitudes, which in turn makes it more challenging to morphologically distinguish between clusters and non-clusters. At magnitude bins of 15, 16, and 17\,mag, the classification purity values of star clusters are 100\%, indicating that no non-clusters are misclassified as star clusters. At magnitude bins of 18, 19, and 20\,mag, the classification purities are 95.83\%, 78.38\%, and 70.36\%, respectively, implying that some non-cluster sources are misclassified as star clusters. Nevertheless, the overall purity of star clusters remains high. At magnitude bin of 21\,mag, the purity is 50.66\%, with the model correctly classifying 305 true star clusters but erroneously labeling 297 non-cluster sources as star clusters. Notably, this magnitude exhibits the highest number of true star clusters as well as the highest number of misclassified non-clusters. At a magnitude bin of 22\,mag, the purity drops drastically to only 21.17\% due to a significant contamination of incorrectly classified non-cluster sources in the test set. At a magnitude bin of 23\,mag, no star clusters are correctly classified since there are only two true star clusters within this magnitude bin in the test set, while the model predicts 30 non-cluster sources as star clusters. In the magnitude bins of 24 and 25\,mag, there are no true star clusters in the test set, resulting in the absence of correctly classified star clusters. However, the model introduces 14 and 5 non-cluster sources, respectively.

\begin{figure}
	\centering
	\includegraphics[width=1.0\linewidth]{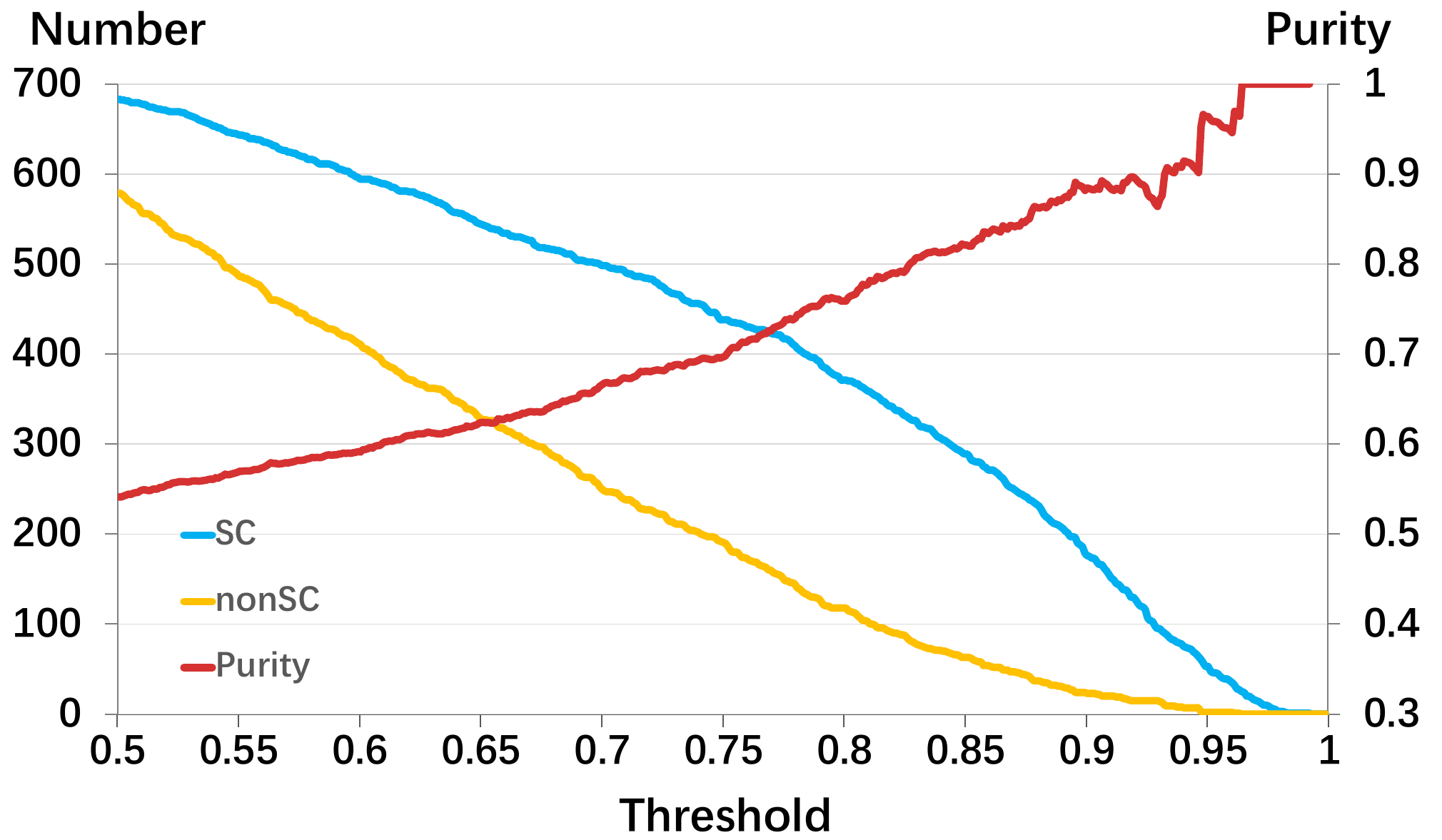}
	\caption{Variations of star cluster classification purity (red curve), the number of correctly classified true star clusters (blue curve), and incorrectly classified non-clusters (yellow curve) for different classification thresholds. These measurements are obtained by evaluating the RestNet-50 model on the test set.}
	\label{thr1}
\end{figure}

Next, we examine the behavior of our trained model on the test set when different classification thresholds are applied. Fig.~\ref{thr1} illustrates the variations in star cluster classification purity and the number of model-predicted star clusters, including the correctly classified true clusters and the misclassified non-clusters, across various classification thresholds. Increasing the classification thresholds of the model implies a more stringent identification criterion for clusters, resulting in higher purity of the prediction results. However, this also leads to a simultaneous decrease in the number of clusters predicted by the model, encompassing both correctly identified true clusters and incorrectly labeled non-clusters. The number of correctly classified star clusters gradually decreases as the classification threshold increases within the range of 0.5 to 0.8. In the threshold range of 0.8 to 0.95, the decrease rate slightly accelerates with increasing thresholds. The overall trend of the number of misclassified non-clusters shows a slight difference compared to the correctly classified true clusters. Within the threshold range of 0.5 to 0.85, the number of star clusters steadily decreases as the threshold increases. In the range of 0.85 to 0.95, the decrease in the number slows down as the threshold increases. The purity of star clusters gradually increases from 55\% to 68\% within the threshold range of 0.5 to 0.75. In the range of 0.75 to 0.92, there are some subtle fluctuations in the purity of star clusters, but overall, it exhibits an increasing trend and improves at a faster rate compared to the previous range.

\subsection{Criteria}

In our study, we focus on selecting star cluster candidates within a projected radius of 30\,kpc from the center of M31. These regions include the PAndAS m207, m210, m212, m228, m233, m249, m254, m265, m285, m248, m257, m266, and m279 fields. The PHAT regions are excluded. This yields a total of 14,136,141 sources from PAndAS photometric data. Using our trained ResNet-50 model, we perform star cluster identification on these objects, resulting in the prediction of 265,856 star cluster candidates. However, their purity is surely to be low. 

In Appendix~\ref{1st_model}, we have discussed the performance of the first model on the test set, emphasizing that both magnitude and classification thresholds are crucial factors influencing the purity of star cluster classifications. The size and color of star clusters have a measurable impact on the model's performance, while the projected distance and distance from the center of CCDs exerts only a limited influence. While pre-filtering candidate sources by size or color could marginally improve accuracy, these cuts would result in the loss of candidates at the extremes of these properties, which are relatively rare but valuable. Therefore, we have excluded the projected distance, size and color of star clusters, as well as their distance from the center of CCDs, from our evaluation criteria. Consequently, we decide to further refine the selection of these 265,856 star cluster candidates based on their magnitude and classification thresholds.

After applying the criteria of a $g$-band magnitude smaller than 21\,mag and a classification threshold greater than 0.669, we have identified a total of 1,299 sources from the 265,856 model-predicted candidates. Through the visual inspection of 1,299 sources conducted by authors BSZ and PJC, we found that the majority of the sources are star cluster candidates.

\section{New M31 Disk Star Cluster Candidates}

Since the first model exhibited optimal performance for sources brighter than 21\,mag, we constructed a specialized training dataset limited to sources with $g$-band magnitudes less than 21\,mag to develop a second ResNet-50 model.  
The training, validation, and test sets were divided consistently with the first method described in Sect.~\ref{sec:Data}. Specifically, the training and validation sets included 879 star clusters, 72 galaxies, 143 H~{\sc ii} regions, 41 PNe, 751 stars, 3 symbiotic stars, and 11,151 remaining non-clusters. The test set contained 490 star clusters and 5,473 non-clusters.
The training and validation data were randomly split in a 7:1 ratio, and data augmentation techniques such as rotations and translations were applied. To maintain a balanced ratio of positive and negative samples, the negative samples were also augmented. A summary of the training data set with $g$-band magnitudes less than 21\,mag is presented in the Table~\ref{dataset2}.

\begin{table}[h]
	\centering
	\caption{Training data set statistics with $g$-band magnitudes less than 21\,mag.}
	\label{dataset2}
	\setlength{\tabcolsep}{2.2mm}{
		\renewcommand{\arraystretch}{1.2}
		\begin{tabular}{cccc}
			\hline
			Class & Training set & Validation Set & Test Set \\
			\hline
			SC & 43,064 &6,160 &490  \\
			nonSC & 43,457 & 6,197 &5,473 \\
			\hline
	\end{tabular}}
\end{table}

\begin{figure}
	\centering
	\includegraphics[width=1.0\linewidth]{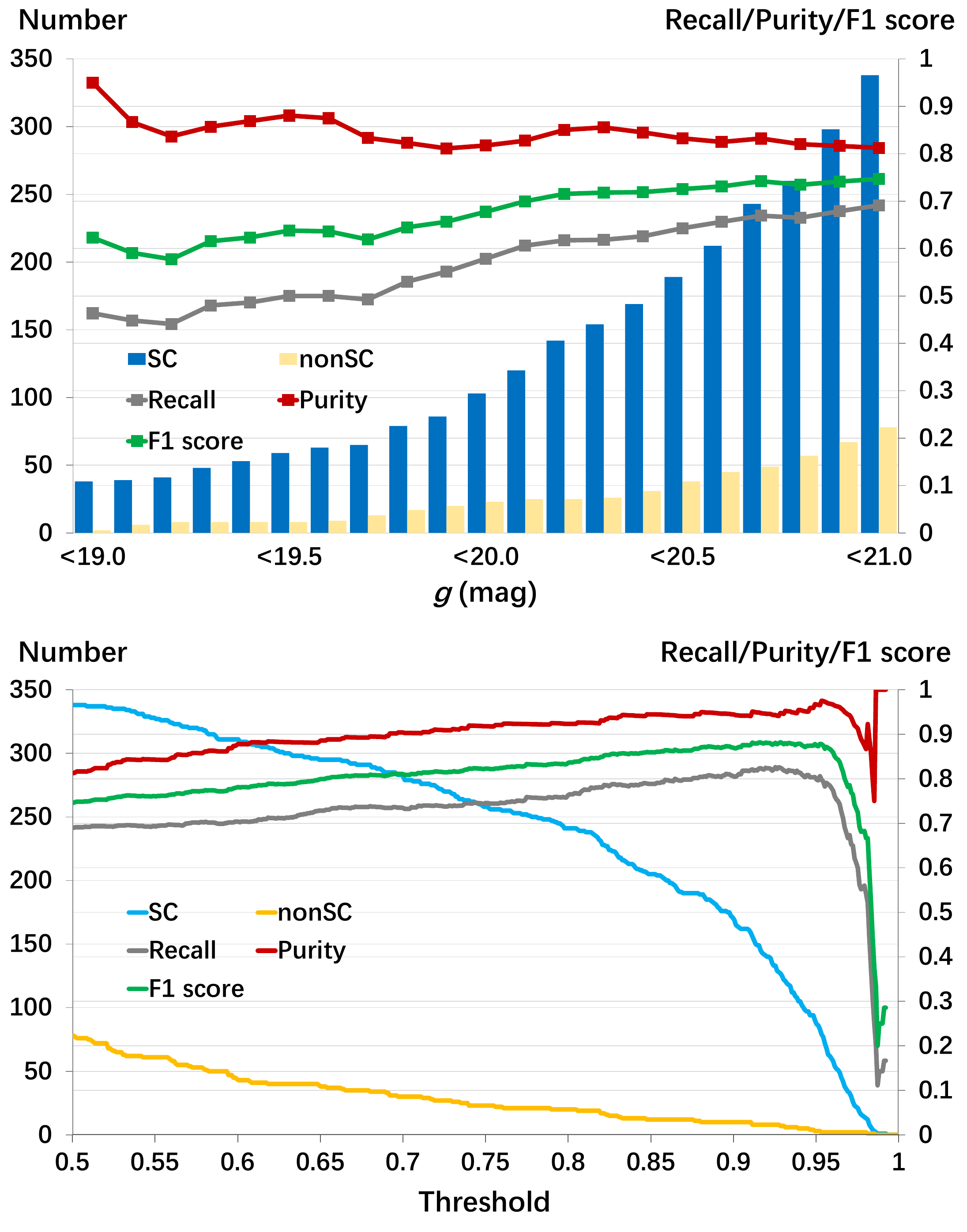}
	\caption{The second model performance varies with star cluster $g$-band magnitudes (upper panel), and threshold (bottom panel).}
	\label{star06_magthr}
\end{figure}

\begin{figure*}
	\centering
	\includegraphics[width=1.0\linewidth]{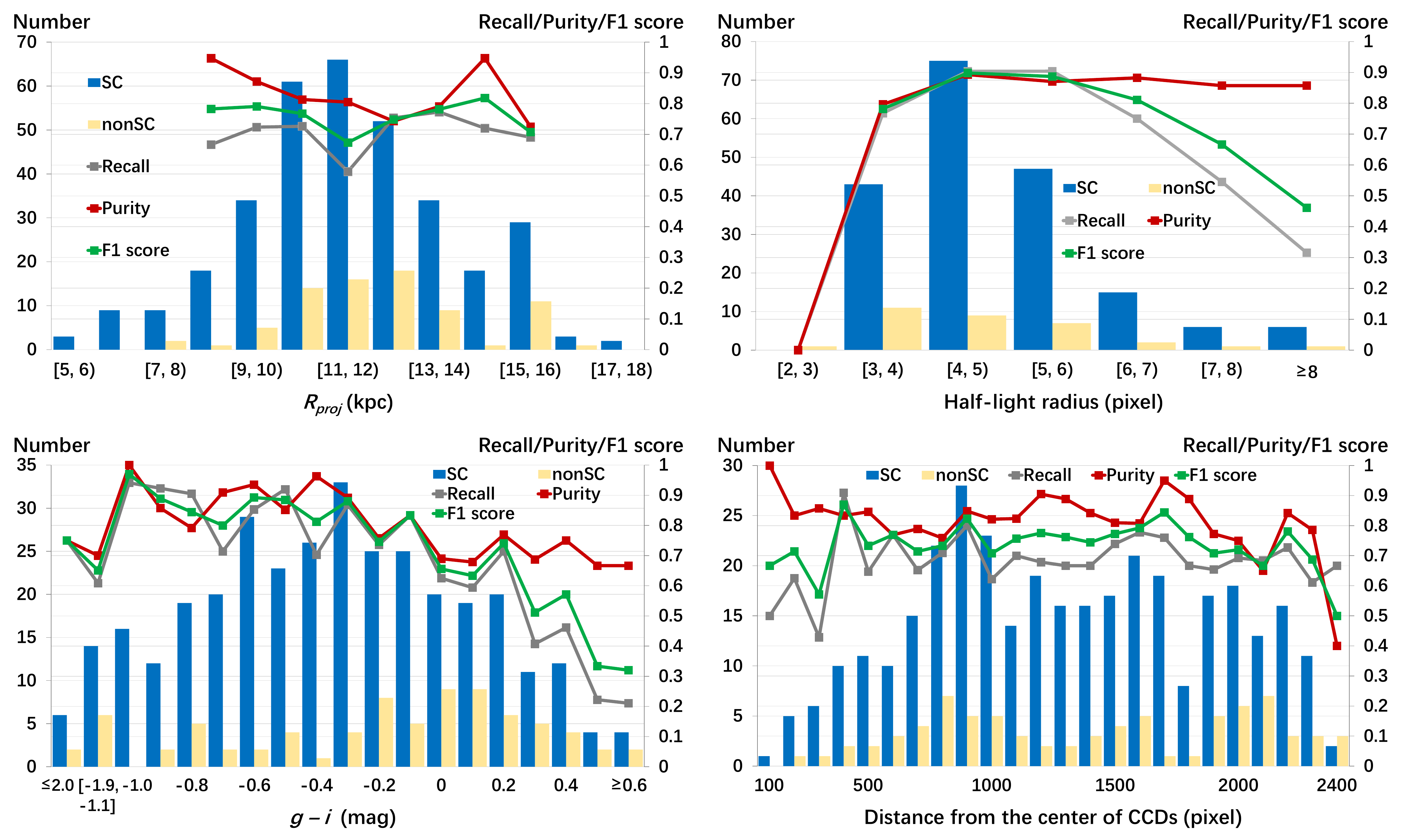}
	\caption{Performance of the second model with respect to various parameters: projected distance ($R_{proj}$) in kpc (upper-left panel), half-light radius in pixels (upper-right panel), ($g$ - $i$) (lower-left panel), and distance from the center of the CCDs in pixels (lower-right panel).}
	\label{star06_pscd}
\end{figure*}

\subsection{Performance of the Second ResNet-50 Model on the Test Set}\label{2nd_model}

After training, the second ResNet-50 model achieved a purity of 81.25\%, a recall of 68.98\%, and an F1 score of 74.61\%. Here, we evaluate the behavior of our trained second model on the test set across various cluster properties, including $g$-band magnitude, classification threshold, projected distance, size, color ($g$ - $i$), and distance from the center of CCDs.

The upper panel of Fig.~\ref{star06_magthr} shows how the performance metrics of the second model vary with the $g$-band magnitudes of the star clusters. The trends are consistent with those of the first model observed in the middle panel of Fig.~\ref{star05_mag_thr} in Appendix~\ref{1st_model}. Purity decreases gradually, while recall and F1 score increase steadily, reaching their optimal values at 21\,mag. The bottom panel of Fig.~\ref{star06_magthr} illustrates the relationship between the performance metrics of the second model and the classification threshold, following a similar trend to the first model observed in the bottom panel of Fig.~\ref{star05_mag_thr}. As the threshold increases, purity generally shows an upward trend with some fluctuations, whereas recall and F1 score first increase and then decline.

Fig.~\ref{star06_pscd} illustrates how the  second model's performance varies with different cluster properties, showing trends consistent with those of the first model presented in Fig.~\ref{star05_pscd} of Appendix~\ref{1st_model}. The projected distance and the distance from the center of the CCDs have only a limited impact on the classification performance. For sources with a half-light radius smaller than 3 pixels or larger than 7 pixels, the model's performance is relatively poor, while it remains relatively stable within the [3, 7) pixel range. In terms of color, the model performs well overall, with only a slight decline in performance observed for extreme color values.

\begin{table*}
%	\centering
	\caption{Classification evaluation for the five subfields of the test set.}
	\label{star06_5test}
	\setlength{\tabcolsep}{3.0mm}{
		\renewcommand{\arraystretch}{1.2}
		\begin{tabular}{ccccccc}
			\hline
			Subfield & $R_{proj}$ (kpc) & Number & Accuracy & Recall & Purity & F1 score \\
			\hline
			R1 &11.3845&  1,015 & 94.88\% & 69.47\% & 74.16\% & 71.74\% \\
			R2 &10.4647&  2,293 & 96.65\% & 69.95\% & 88.82\% & 78.26\% \\
			R3 &10.3948&  1,055 & 95.07\% & 62.35\% & 72.60\% & 67.09\% \\
			R4 &12.8544&  820   & 96.71\% & 75.95\% & 88.24\% & 81.63\% \\
			R5 &16.5834&  834   & 97.12\% & 63.16\% & 70.59\% & 66.67\% \\	
			\hline
	\end{tabular}}
\end{table*}

\begin{table*}
	\centering
	\caption{Simbad matched FP objects.}
	\vspace{-4mm}
	\label{fp_cls}
	\setlength{\tabcolsep}{1.1mm}{
		\renewcommand{\arraystretch}{1.35}
		\begin{tabular}{c|cccccccccc}
			\hline
			Type&Cl*&GlC&HII&Galaxy&AGN/AG?&V*&Cepheid/ClassicalCep&EmObj&WR*&RedSG\_Candidate\\
			\hline
			Number&1&2&8&3&2&2&3&1&1&1\\
			\hline
	\end{tabular}}
\end{table*}

We then analyzed the performance of the second model across five subfields within the test region. The evaluation metrics, summarized in Table~\ref{star06_5test}, exhibit similar trends to those of the first model (see Table~\ref{5test} in Appendix~\ref{1st_model}) but show an overall improvement in performance.

\begin{figure}
	\centering
	\includegraphics[width=0.85\linewidth]{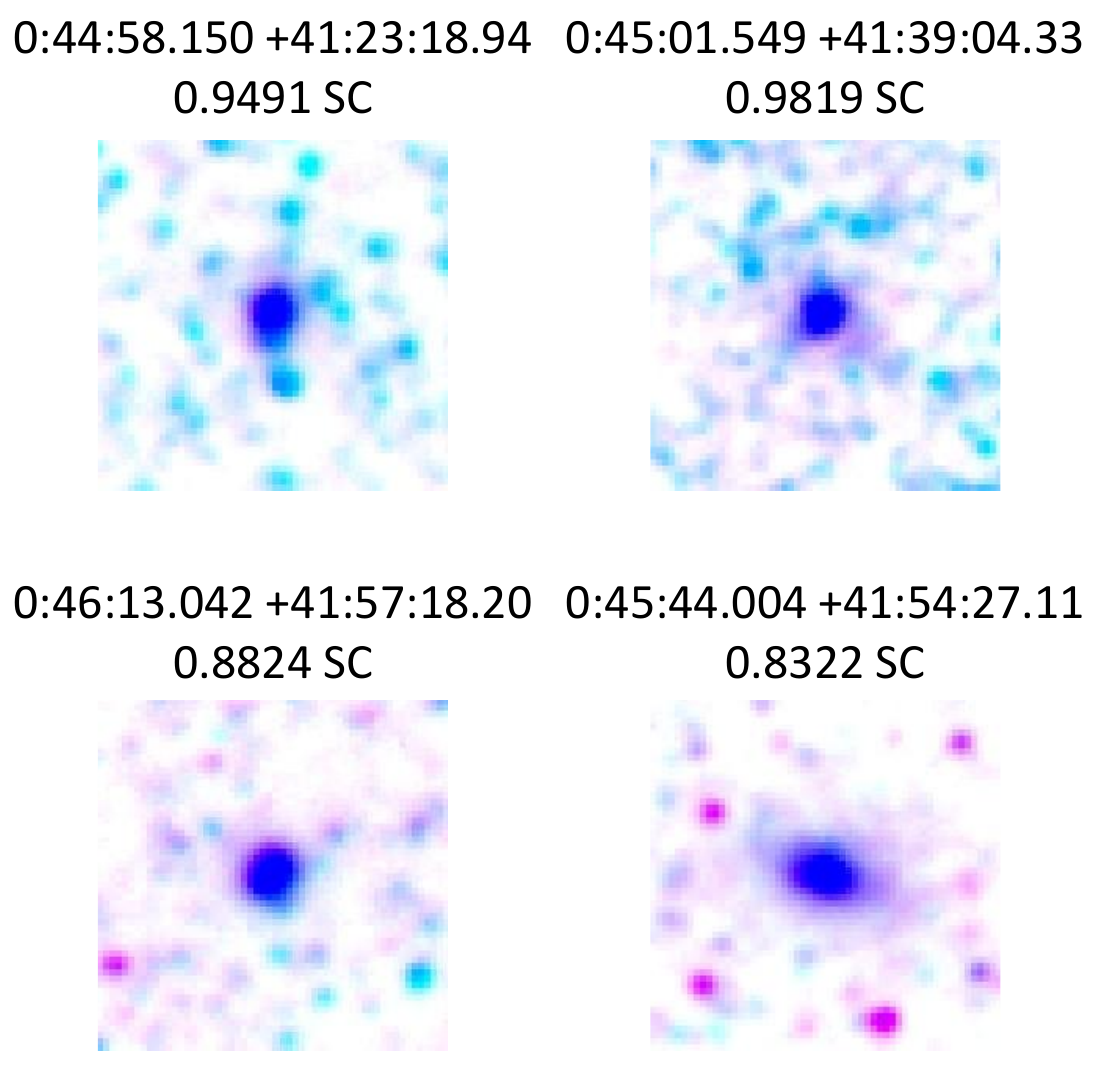}
	\caption{PAndAS $g$ and $i$-band synthesized images of four example FP objects. The coordinates (RA, Dec) from the PAndAS catalog, confidence and corresponding class provided by the ResNet-50 model are labeled above each panel.}
	\label{star06_fp}
\end{figure}

Finally, we examine the sources of misclassified objects predicted by our second model on the test set, which include both false positives (FPs) and false negatives (FNs). We begin by analyzing the sources of FPs. To achieve this, we performed a cross-match between the 78 FP objects and the SIMBAD database, using a matching radius of 1". This process yielded 32 matched records. Among these, eight objects were identified as young disk star clusters, as reported in the studies by \cite{johnson2012phat} and \cite{johnson2015phat}, which were mismatched with the PAndAS catalog in this work. The types and numbers of the remaining 24 matched objects are summarized in Table~\ref{fp_cls}. Notably, three star clusters are reported, two of which are classified as old globular clusters. The remaining sources are primarily classified as galaxies (3) and H~{\sc ii} regions (8). These contaminants likely arise due to the visual resemblance between the morphological features of galaxies and H~{\sc ii} regions and those of star clusters. Additionally, the analysis identified three objects classified as Cepheids or Classical Cepheids and two as variable stars, both of which contribute to the contamination. Based on cross-referencing with the SIMBAD and our visual inspection, the majority of FP objects exhibit characteristics of extended sources. The primary sources of contamination are background galaxies and H~{\sc ii} regions. Fig.~\ref{star06_fp} presents PAndAS images of several representative FP objects. The morphology of these FP objects—where the image in the bottom-right panel depicts a galaxy—closely resembles that of star clusters, making them particularly challenging to distinguish.

\begin{figure}
	\centering
	\includegraphics[width=0.85\linewidth]{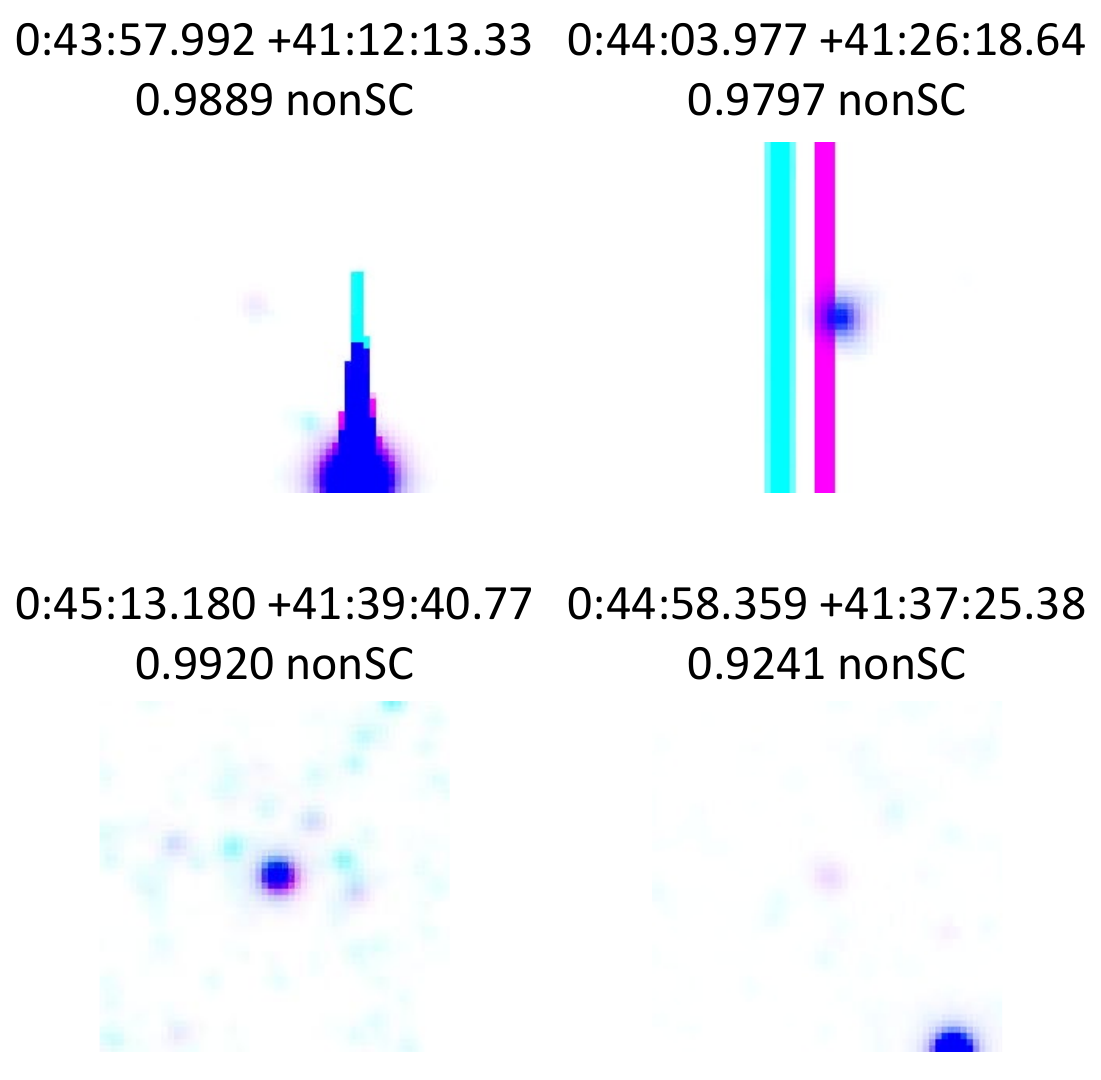}
	\caption{Same as Fig.~\ref{star06_fp}, PAndAS $g$ and $i$-band synthesized images of four example FN objects.}
	\label{star06_fn}
\end{figure}

For the FN objects, we examined their PAndAS images to identify the causes of misclassification. Many of these objects were misclassified due to poor image quality. Specifically, they either appear faint, are affected by overexposed neighboring sources, or have very small sizes, all of which make them difficult to classify.
Fig.~\ref{star06_fn} presents four representative examples. The upper panels show two star clusters that appear faint due to the presence of nearby overexposed sources, which significantly degrade their imaging quality. The star cluster in the bottom-left panel is a very small source, further complicating its classification. Additionally, the star cluster shown in the bottom-right panel has a combination of faintness and morphology that makes its classification challenging, even upon visual inspection. These examples underscore the difficulties posed by poor image quality, faintness, and small sizes in the classification of star clusters.

\subsection{Visual Inspection}

The bottom panel of Fig.~\ref{star06_magthr} indicates that the  second model achieves its highest performance at a classification threshold of 0.929. However, adopting such a high threshold would eliminate a significant number of star clusters. To ensure consistency with the first model, we retain the classification threshold of 0.669. This value balances cluster retention and purity, aligning with our study’s requirements.

\begin{figure}[h]
	\centering
	\includegraphics[width=0.82\linewidth]{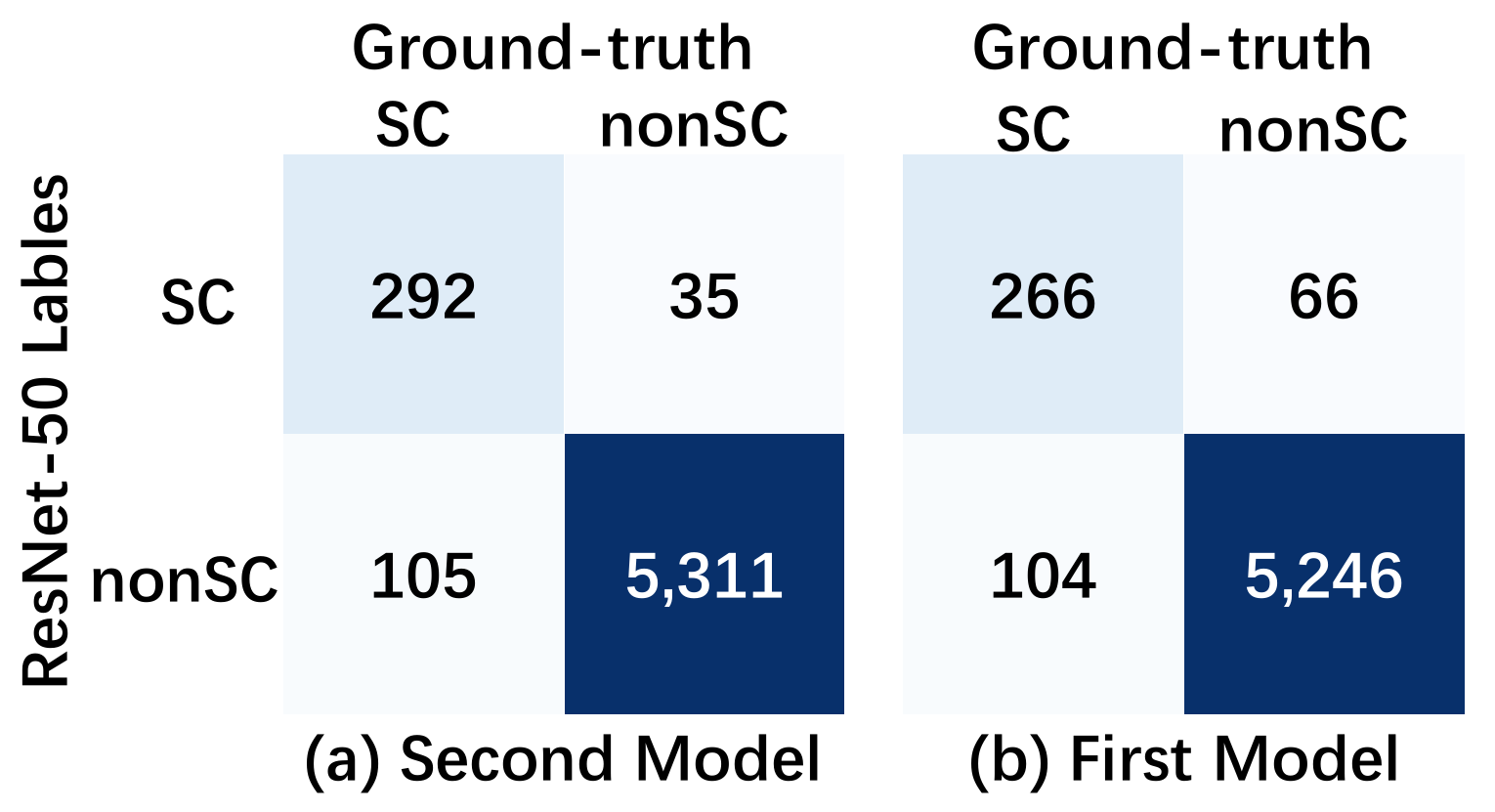}
	\caption{Confusion matrices for two models at classification threshold of 0.669. The left panel corresponds to the second model, while the right panel represents the first model.}
	\label{second_cm}
\end{figure}

When the classification threshold is set to 0.669, the second model achieves a purity of 89.30\%, a recall of 73.55\%, and an F1 score of 80.66\%. These results show consistent improvements over the first model across all metrics. The confusion matrices for both models at this threshold are presented in Fig.~\ref{second_cm}, where the left panel corresponds to the second model and the right panel represents the first model. In Fig.~\ref{second_cm}, the second model correctly identifies 292 star clusters and misclassifies 35 false positives (nonSC as SC), while the first model identifies 266 star clusters and misclassifies 66 false positives. This indicates that the second model achieves higher true positive rates and significantly reduces false positive rates compared to the first model. Additionally, the second model maintains a comparable performance on true negatives (5,311 vs. 5,246). 
These results suggest that the second model provides more reliable classification, especially for identifying star clusters, while maintaining a low false positive rate. 

Using the second model and threshold, we identified 2,605 star cluster candidates from 108,076 images with $g$-band magnitudes less than 21\,mag. We cross-matched 2,605 sources generated by the second model with 1,299 sources from the first model and identified 1,110 common sources. 

Although our ultimate goal is to automate the identification process and eliminate the reliance on visual inspection, this study serves as an exploratory piece of experimental and methodological research. For this reason, we conducted visual inspections to validate or refute the results produced by our automated methods. Prior to conducting the visual inspection, we removed duplicate candidate sources, resulting in a final set of 2,228 unique independent candidate sources. The visual inspection scoring system is as follows: 2 points for confirmed star clusters, 1 point for potential star clusters, and 0 points for non-clusters. Two experienced authors, BSZ and PJC, manually inspected a total of 2,228 sources. The number of sources receiving scores of 0, 1, 2, 3, and 4 points were 499, 238, 434, 389, and 668, respectively. Through visual inspection, we identified four main types of contamination in the candidate sources obtained by the model: poor-quality images, background galaxies, binary stars, and faint sources. We select sources with scores of 3 and 4, resulting in a final catalog of 1,057 M31 star cluster candidates, including 745 newly identifications.

\begin{figure}
	\centering
	\includegraphics[width=1.0\linewidth]{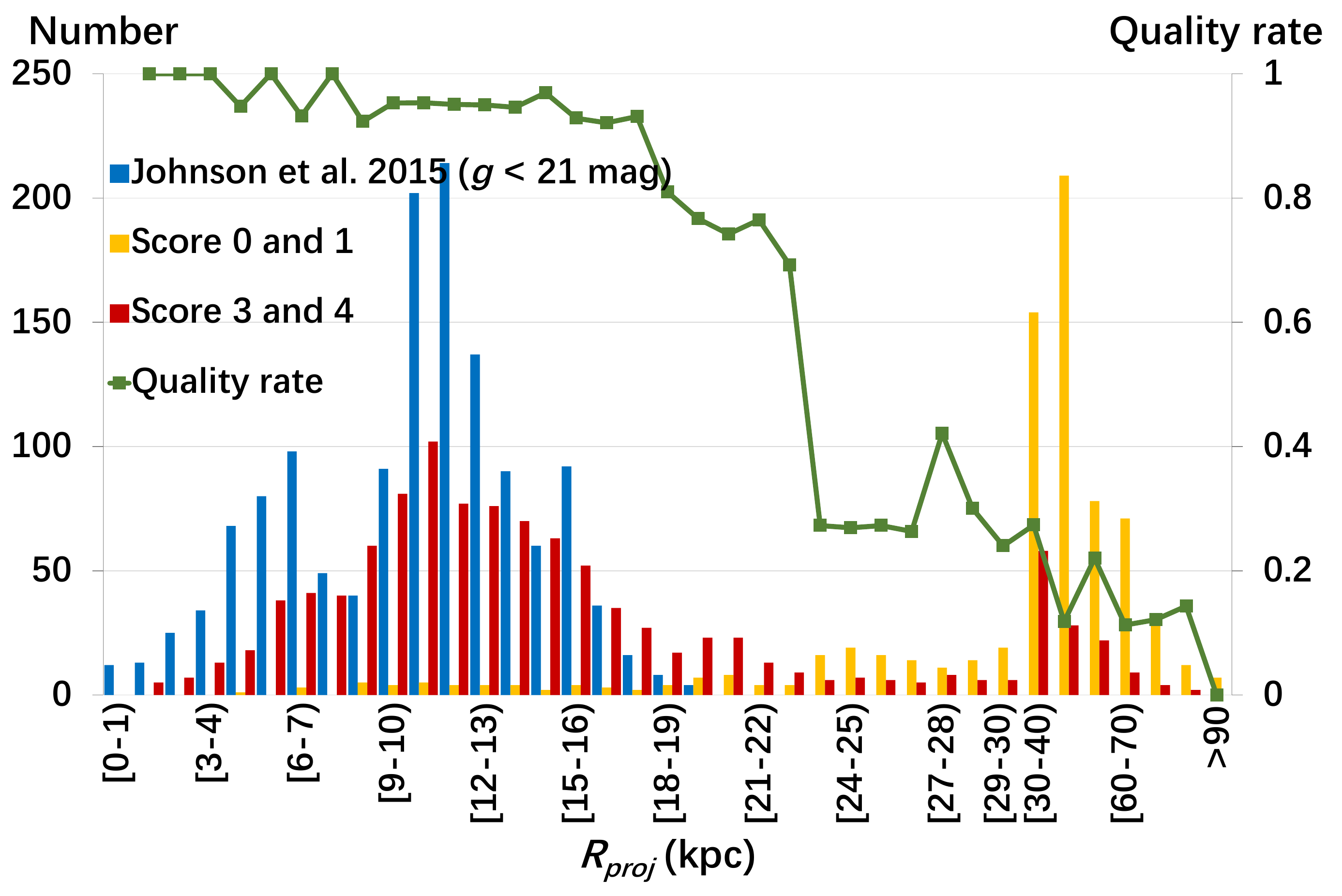}
	\caption{Variations of the quality rate (green curve), the number of star clusters from \cite{johnson2015phat} (blue bars), the number of sources with scores 0 and 1 (yellow bars), and the number of sources with scores 3 and 4 (red bars) along with the projected distances. We note that the blue bars represent the number of best-matching entries from the PAndAS photometry catalog, rather than the number of PHAT star clusters.
	The data is divided into bins based on projected distance, with a bin width of 1\,kpc for distances from 0 to 30\,kpc and 10\,kpc for distances greater than or equal to 30\,kpc.}
	\label{qr}
\end{figure}

\begin{figure}
	\centering
	\includegraphics[width=0.85\linewidth]{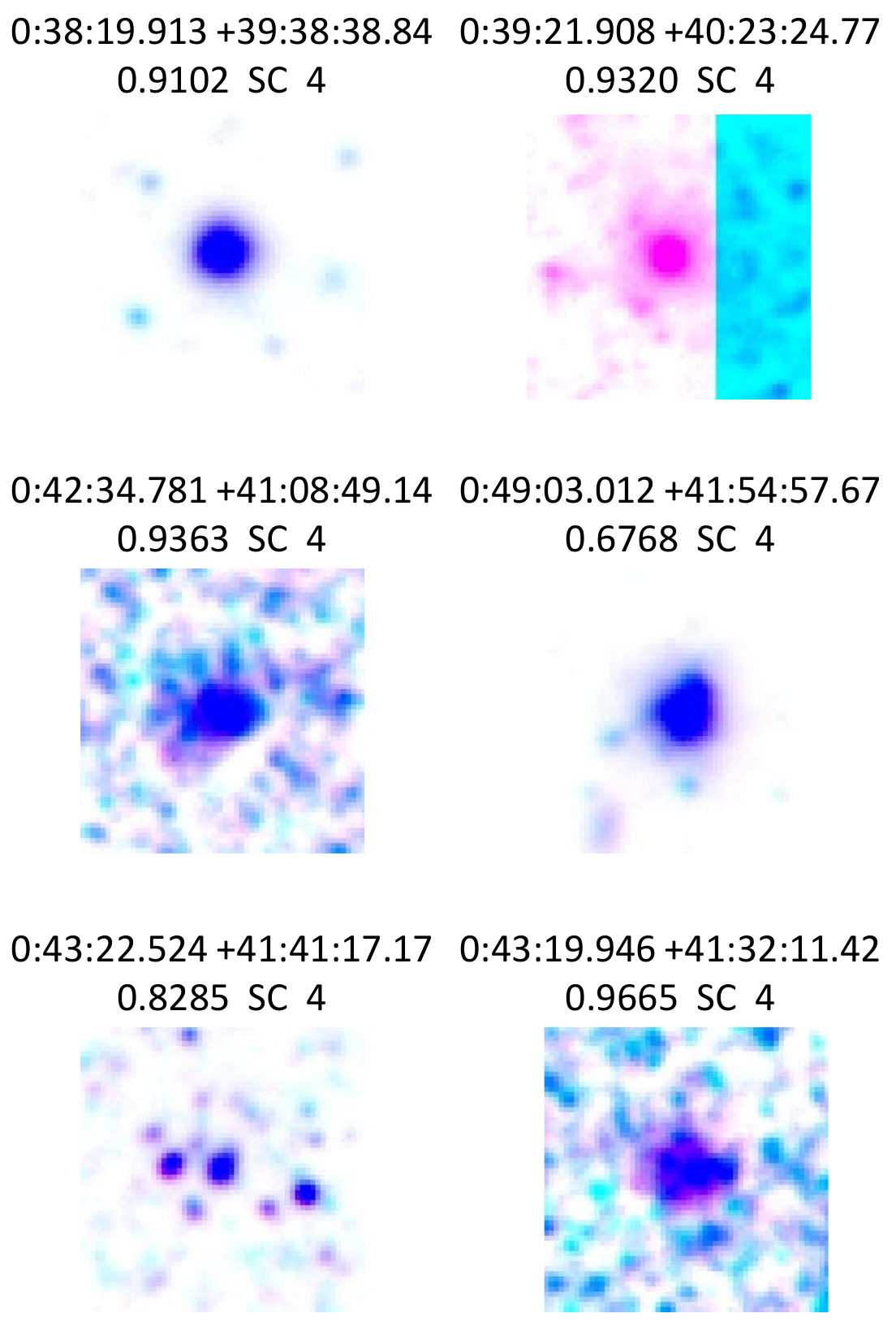}
	\caption{PAndAS synthesized  $g$- and $i$-band images of the six example newly discovered M31 star cluster candidates in the this work. The coordinates (RA, Dec) from the PAndAS catalog, confidence, corresponding class provided by the ResNet-50 model and visual inspection score are labeled above each panel.}
	\label{6sc}
\end{figure}

Fig.~\ref{qr} illustrates the relationship between projected distance and several key parameters, including the quality rate and the number of star clusters. The x-axis represents the projected distance, divided into bins of 1\,kpc for the range 0 – 30\,kpc and 10\,kpc for distances greater than 30\,kpc. The left y-axis corresponds to the number of clusters, whereas the right y-axis represents the quality rate. The quality rate, defined as the ratio of the number of clusters with high-quality sources (scores 3 and 4) to the number of clusters with both high- and low-quality sources (scores 3, 4, 0, and 1), shows a clear decline with increasing projected distance. 
Within 20\,kpc, there are sufficient training samples, resulting in a larger number of sources with scores 3 and 4 and fewer sources with scores 0 and 1. Consequently, the quality rate remains relatively high (above 80\%), indicating that most candidates in this range are of high quality. Beyond 20\,kpc, the lack of sufficient training samples leads to a decrease in the number of sources with scores 3 and 4 and an increase in the number of sources with scores 0 and 1. As a result, the quality rate drops significantly, falling below 30\% at distances greater than 30\,kpc.

\begin{figure}
	\centering
	\includegraphics[width=0.99\linewidth]{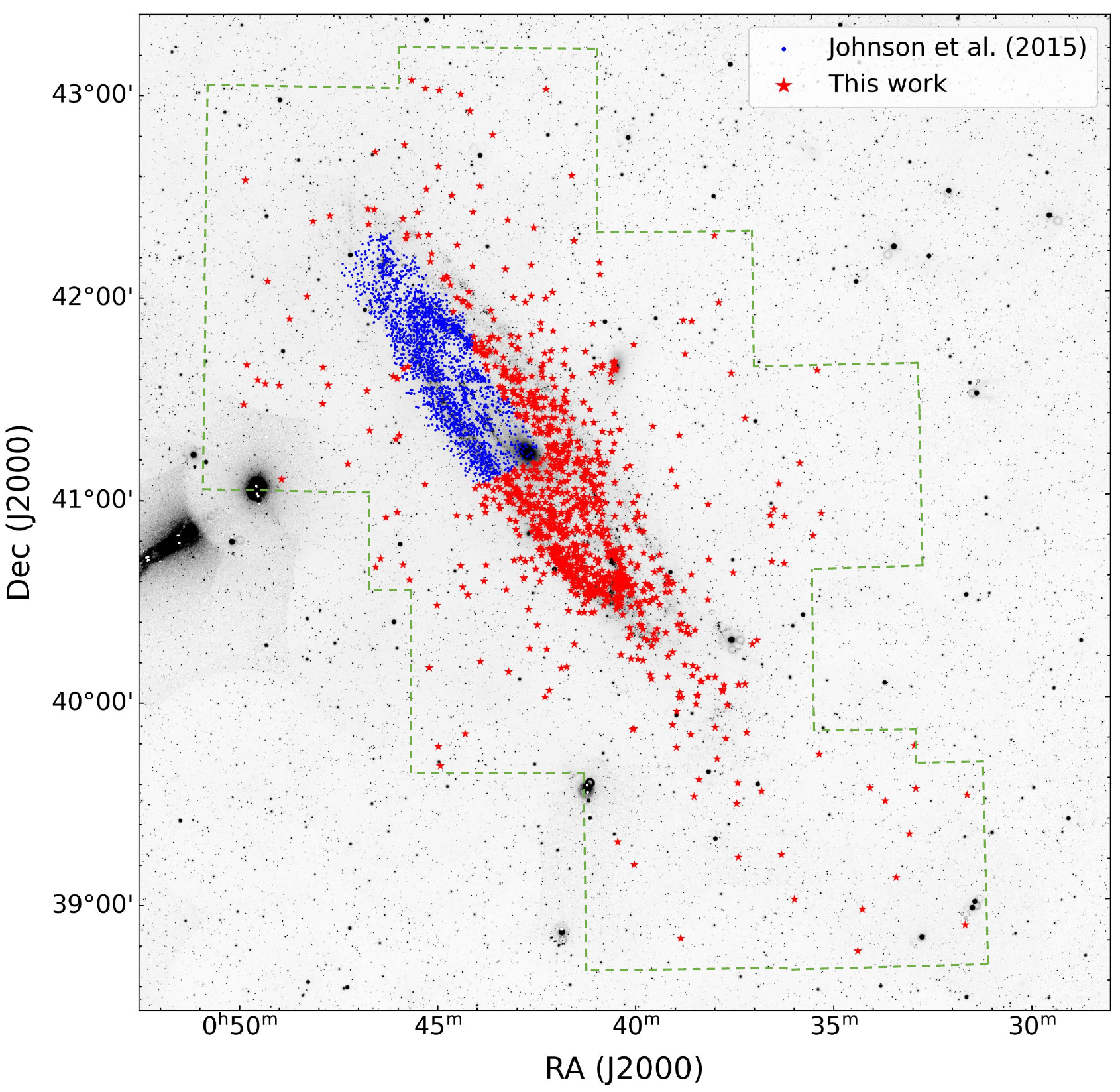}
	\caption{Spatial distribution of the 1057 M31 star cluster candidates (red stars) and the PHAT star clusters from \cite{johnson2015phat} (blue dots). We note that the blue points are the best matching entries from the PAndAS photometry catalog, and not the original PHAT coordinates. The region enclosed by the green dashed line demarcates the area of the sky scrutinized in the PAndAS study for this research. The background image is taken from the GALEX NUV observation.}
	\label{spatial}
\end{figure}

\subsection{Catalog of Our Identified Star Clusters}

\begin{table*}
	\centering
	\caption{Catalog of the newly identified 1,057 star cluster candidates in M31. The full table for the 1,057 star cluster candidates can be accessed on the website at \url{https://nadc.china-vo.org/res/r101475/}}
	\label{sc_candt}
	\setlength{\tabcolsep}{0.8mm}{
		\renewcommand{\arraystretch}{1.2}
			\begin{tabular}{cccccccccccc}
				\hline
				Name&RA&Dec&g&$\sigma_g$&i& $\sigma_i$&PAndAS&$R_{proj}$&Confidence&Score&Note\\
				&(J2000)&(J2000)&(mag)&(mag)&(mag)&(mag)&field ID&(kpc)&&&\\
				\hline
				Candidate1& 0:31:27.527& +39:32:21.78& 18.636& 0.001& 17.864& 0.001& 207& 53.79& 0.904& 4& PAndAS-21\_from\_H14\\
				Candidate2& 0:31:37.216& +38:53:53.27& 19.797& 0.002& 18.389& 0.001& 207& 45.49& 0.8474& 3&\\ 
				Candidate3& 0:32:46.534& +39:34:40.44& 13.559& 0.001& 12.744& 0.001& 207& 44.42& 0.7867& 4& G001-MII\_from\_C09,\\
				&&&&&&&&&&&G001\_from\_RBC\\
				Candidate4& 0:33:33.787& +39:31:18.83& 15.398& 0.001& 14.46& 0.001&  207& 38.44& 0.8444& 4& G002-MIII\_from\_C09,\\ 
				&&&&&&&&&&&G002\_from\_RBC\\
				Candidate5& 0:34:21.586& +38:46:59.67& 20.628& 0.003& 19.969& 0.004& 207& 41.95& 0.7004& 3&\\ 
				Candidate6& 0:36:14.995& +39:16:08.50& 20.095& 0.002& 19.513& 0.003& 210& 34.41& 0.9441& 3&\\
				…&	…&	…&	…&	…&	…&	…&	…&	…&…&…&…\\
\hline
\end{tabular}}
\end{table*}

Table~\ref{sc_candt} presents the coordinates, $g$- and $i$-band magnitudes, PAndAS field ID, projected distance to the center of M31, the classification confidence, visual inspection score and the ‘Note’ column of these 1,057 star cluster candidates. The ‘Note’ column in Table~\ref{sc_candt} indicates that the source has already been found in any of the previous works. 
The $g$- and $i$-band magnitudes from PAndAS photometry are aperture magnitudes. In principle, the aperture magnitudes are robust for extended sources such as clusters. However, we acknowledge that for some semi-resolved sources, the photometry may not be accurate. Nevertheless, our dataset comprises a total of 3,202 star clusters (There are a total of 2,142 star clusters in the training and validation sets, and 1,060 star clusters in the test set). Among these, 2,798 are flagged as extended sources in both the $g$ and $i$ bands in the PAndAS catalog. Additionally, 286 are flagged as extended sources in either the $g$ or $i$ band. The remaining 118 clusters are flagged as either point sources or noise. 
In Fig.~\ref{6sc}, we display the $g$- and $i$-band synthesized images of six example newly discovered star cluster candidates in M31.

\begin{figure}
	\centering
	\includegraphics[width=0.95\linewidth]{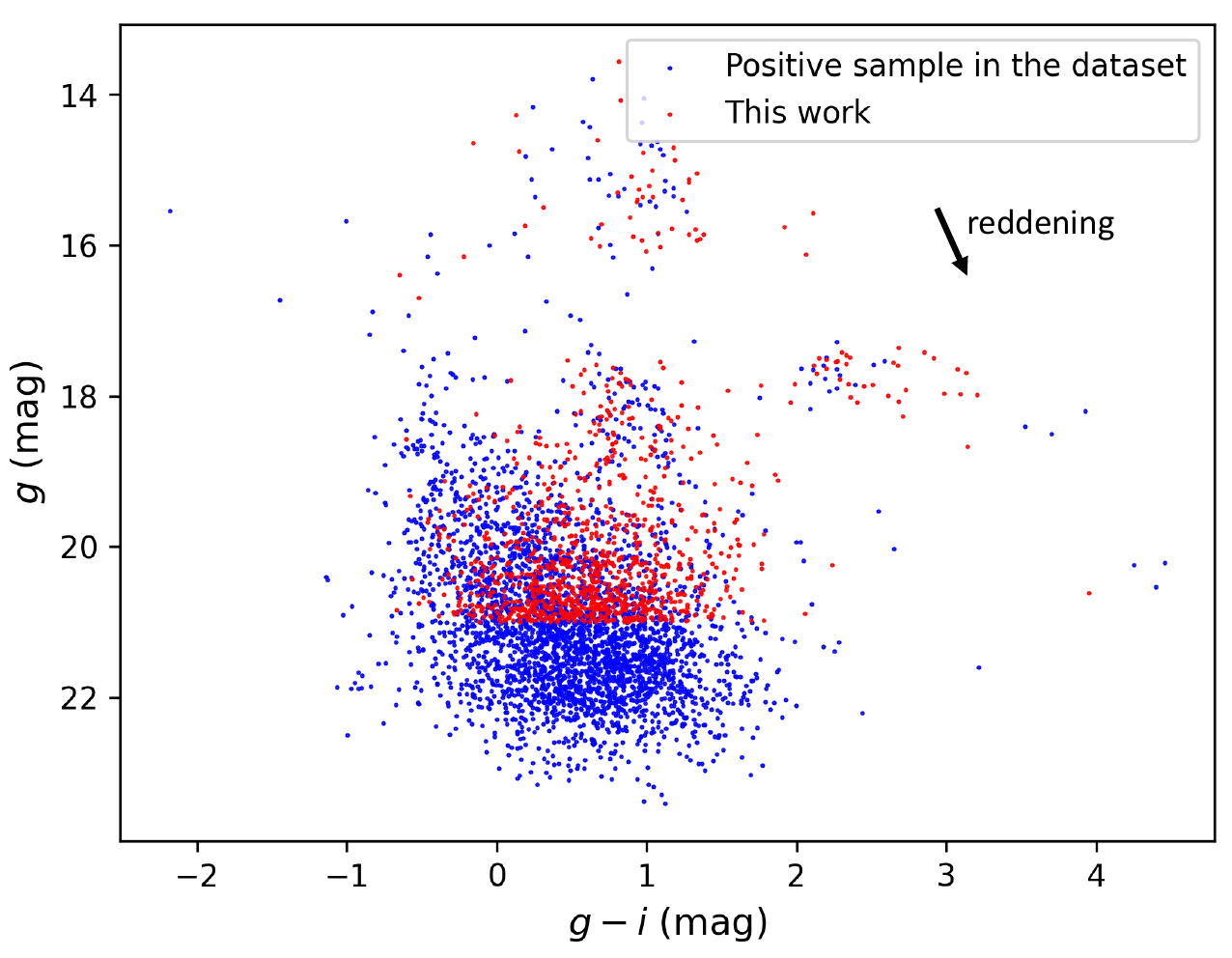}
	\caption{Color-magnitude diagram of the newly identified M31 star cluster candidates (red dots) and the positive sample from the dataset in Sect. 2 (blue dots). The $g$ and $i$-band photometry for these star clusters is derived from the PAndAS catalog, rather than the original PHAT photometry. The extinction vector, denoted by a black arrow, corresponds to an extinction value of $A_V=1$. This vector is drawn based on the extinction law from \cite{cardelli1989relationship} and \cite {o1994rnu}.}
	\label{cmd}
\end{figure}

The spatial distribution of the 1,057 identified star cluster candidates is depicted in Fig.~\ref{spatial}. Similar to the PHAT clusters from \cite{johnson2015phat}, a significant portion of these discovered candidates is situated within the spiral arms of M31, suggesting the reliability of our identification methodology for young disk clusters. Notably, there exist some star cluster candidates located at projected distance larger than 30\,kpc. We focused our inspection on sources with a projected distance greater than 30\,kpc, the visual inspection score is equal to 4 points, and those not previously identified in previous work, resulting in a total of 9 such sources. These candidates stand out as strong detections, warranting further investigation such as spectroscopic observations.

\begin{figure}
	\centering
	\includegraphics[width=0.95\linewidth]{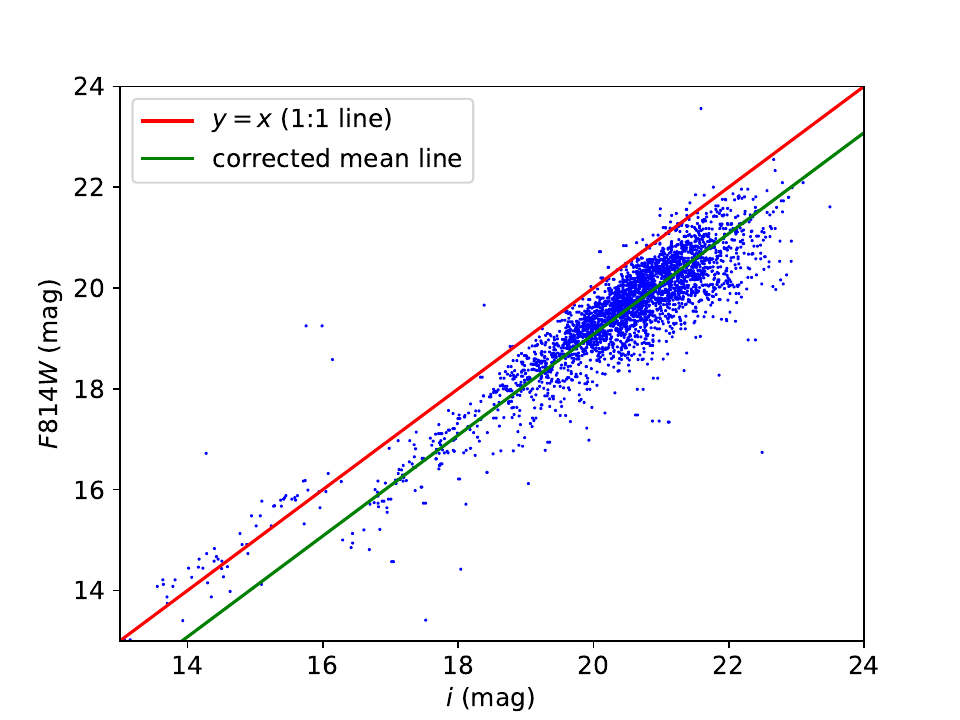}
	\caption{Comparison between PAndAS $i$-band magnitudes and the PHAT $F814W$ magnitudes for all clusters in the positive samples of dataset. The red line is corresponding to the equation $y = x$, and the green line represents the line corrected for the mean of the magnitude difference distribution.}
	\label{f814w_i}
\end{figure}

We have plotted the color-magnitude diagram (CMD) of the new star cluster candidates, as shown in Fig.~\ref{cmd}. The CMD was constructed using PAndAS $g$- and $i$-band magnitudes. These newly identified star cluster candidates exhibit comparable colors and magnitudes to those from the PHAT survey by \cite{johnson2015phat}. We have compared the PAndAS $i$-band magnitudes to the PHAT $F814W$ magnitudes for all clusters in the positive samples, as shown in Fig.~\ref{f814w_i}. The red line ($y = x$) is parallel to the green line, indicating that the PAndAS photometry aligns well with the PHAT photometry. The offset between the two lines suggests a systematic difference between the two datasets of different filters. Most clusters are located around the green line, with only a few outliers exhibiting very large magnitude differences.

\begin{table}[h]
%	\centering
	\caption{The catalog is cross-matched to other existing work.}
	\label{cross-match}
	\setlength{\tabcolsep}{0.5mm}{
		\renewcommand{\arraystretch}{1.35}
		\begin{tabular}{cc}
			\hline
			Source&Number of common stars\\
			\hline
			\cite{wang2022identification}&23\\
			\cite{wang2023searching}&5\\
			\cite{caldwell2008star}&265\\
			\cite{caldwell2011star}&170\\
			\cite{huxor2014outer}&1\\
			Revised Bologna Catalog\footnote{\url{https://cdsarc.cds.unistra.fr/viz-bin/cat/V/143}}&271\\
			\hline
	\end{tabular}}
\end{table}

We have compared our candidates with those from \cite{wang2022identification}. Upon visual inspection, \cite{wang2022identification} selected 117 M31 star cluster candidates. Within our target region, there are 28 sources from Wang et al.'s catalog that possess a $g$-band magnitude smaller than 21\,mag. Out of these, 23 sources are included in our final catalog. This outcome suggests that our model and threshold settings effectively identify trustworthy star cluster candidates. 
In addition, we also cross-matched the catalog with other previous work, and the matching results are shown in Table~\ref{cross-match}. Among them, 5 sources were matched with \cite{wang2023searching} catalog, 265 sources were matched with \cite{caldwell2008star} catalog, 170 sources were matched with \cite{caldwell2011star} catalog, 271 sources were matched with Revised Bologna Catalog, and one source were matched with \cite{huxor2014outer} catalog. 
It is noteworthy that the data used to train and test our model consist of young star clusters located in the disk of M31, whereas the \cite{huxor2014outer} catalog are old globular clusters in the halo. This fundamental difference in the nature of the data may explain the discrepancy. Additionally, the \cite{huxor2014outer} catalog includes only 10 sources within our search range.

\section{Conclusion}

In this work, we have proposed a novel automated approach to identify M31 new disk clusters from large amounts of photometric images, achieving high purity. Using the PAndAS images and catalogs from PHAT surveys, we have trained four deep learning models. Through evaluation on the test set, we determine that the deep learning residual network, ResNet-50, exhibits superior performance. However, on real and unbalanced data, the trained model demonstrates relatively low precision, likely due to the significant imbalance between clusters and non-clusters. To ensure a high-purity sample of cluster candidates, we conduct further analysis on the properties of the model-predicted cluster candidates, including magnitude and model classification thresholds. 
By setting the limiting magnitude to 21\,mag and the classification threshold to 0.669, we successfully obtain new cluster candidates with 80\% purity. To further enhance the purity of the catalog, we focused on training data set with $g$-band magnitudes less than 21\,mag to train a second ResNet-50 model. At classification threshold of 0.669, the second model achieved a purity of 89.30\%, a recall of 73.55\%, and an F1 score of 80.66\%, showing consistent improvements over the first model across all metrics. We applied the second model and identified 2,228 independent star cluster candidates. We conducted visual inspections to validate or refute the results produced by our automated methods. Ultimately, 1,057 star cluster candidates were obtained through visual inspection, including 745 newly identified star cluster candidates.

In the future, we intend to utilize this method to the data from the China Space Survey Telescope Optical Survey (CSST-OS; \citealp{zhan2011consideration, zhan2018overview}). This will allow us to effectively and robustly search for new cluster candidates in nearby galaxies.

\section*{Acknowledgments}

We want to thank the anonymous referee for detailed and constructive comments that improved the manuscript significantly. We thank Ms Lin Zhang, Ms Xingzhu Zou and Mr. Xianhui Meng for useful discussions. This work is partially supported by the National Key Research and Development Program of China No. 2019YFA0405500, National Natural Science Foundation of China 12173034, 12322304 and 11803029, Yunnan Provincial Education Department Scientific Research Fund No. 2024Y039. We acknowledge the science research grants from the China Manned Space Project with no. CMS-CSST-2021-A09, CMS-CSST-2021-A08, and CMS-CSST-2021-B03. 

The numerical computations were conducted on the Yunnan University Astronomy Supercomputer.

\appendix
\vspace{-8mm}
\section{Performance of the First Model on the Test Set}\label{1st_model}

\begin{figure}[h]
	\centering
	\includegraphics[width=0.5\linewidth]{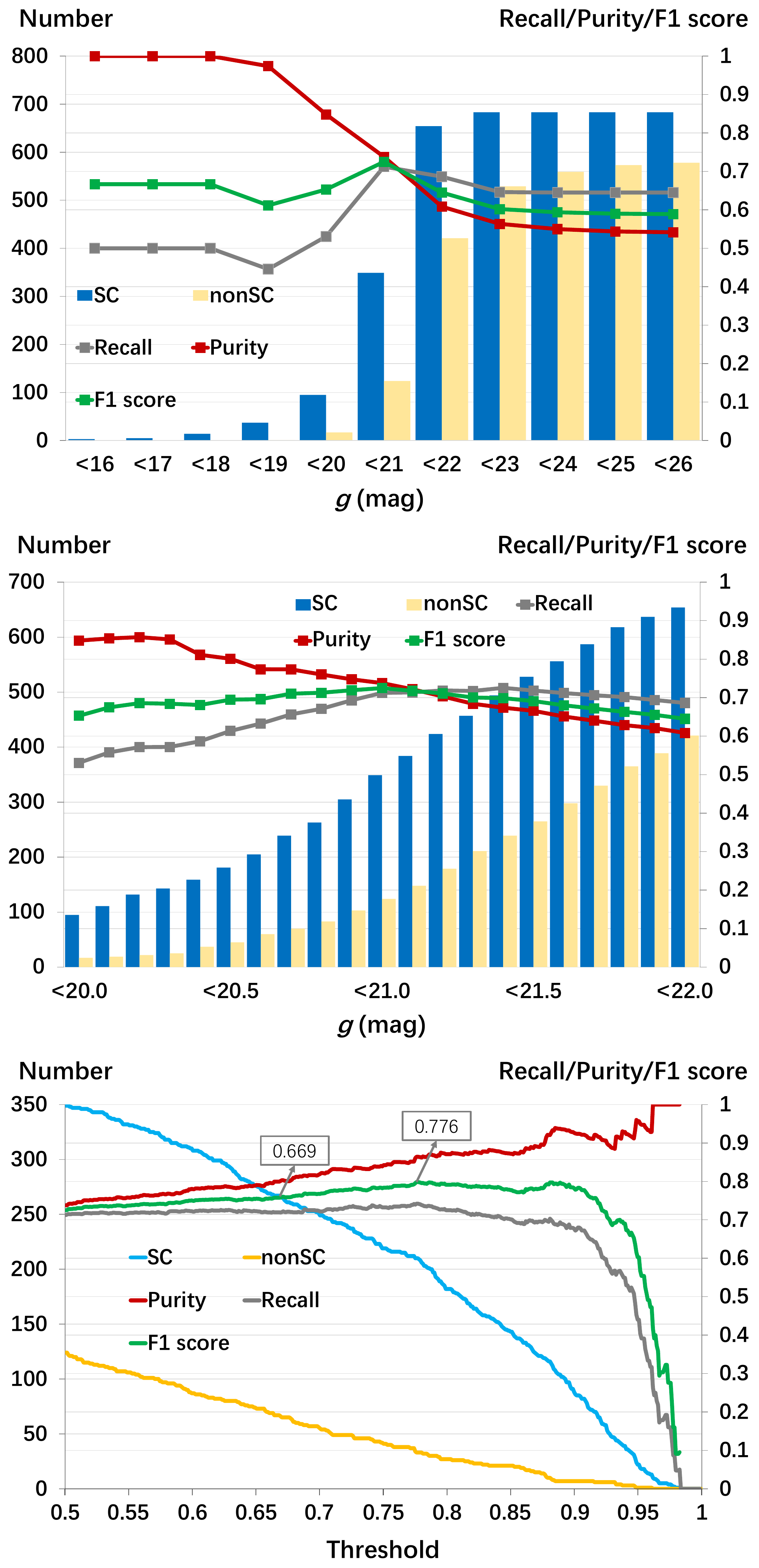}
	\caption{The first model performance varies with star cluster $g$-band magnitudes (upper and middle panels), and threshold (bottom panel).}
	\label{star05_mag_thr}
\end{figure}

We present the variations of recall, purity, F1 score, and the number of model-predicted clusters across various magnitude bins for sources with $g$-band magnitudes smaller than 16\,-\,26\,mag on the upper panel of Fig.~\ref{star05_mag_thr}. These measurements are obtained by evaluating the trained RestNet-50 model on the test set. As depicted in Fig.~\ref{star05_mag_thr}, our classification model achieves the highest F1 score, for the bin of $g$-band magnitudes smaller than 21\,mag. However, binning the sources with 1\,mag is somewhat rough, using smaller increments may produce better F1 score. According to the upper panel, we can conclude that the best F1 score should be between 20\,-\,22\,mag. We bin for sources with $g$-band magnitudes smaller than 20\,-\,22\,mag using 0.1\,mag, as shown in the middle panel Fig.~\ref{star05_mag_thr}, our classification model achieves the highest F1 score (72.48\%) and recall (71.22\%), along with a purity of 73.78\%, for the bin of $g$-band magnitudes smaller than 21\,mag. This indicates optimal performance within this magnitude range. Consequently, we restrict the magnitude of the candidate star clusters to be less than 21\,mag.

We proceed by analyzing the performance of our trained model on sources with $g$-band magnitudes smaller than 21\,mag in the test set, using different classification thresholds. The variations in star cluster classification purity, recall, F1 score, and the number of model-predicted star clusters are illustrated at the bottom panel of Fig.~\ref{star05_mag_thr}. From the Figure, it can be observed that the maximum F1 score of 79.85\% is achieved when the classification threshold is set to 0.776, indicating the best overall performance of the model. However, at this threshold, the number of correctly predicted true star clusters and misclassified non-clusters are only 210 and 33, respectively. This threshold leads to the exclusion of a significant number of sources. To strike a balance between performance and source preservation, we further optimize our criteria by setting the classification threshold to 0.669, which leads to  a overall purity of 80\%. At this threshold, the F1 score and recall are 75.78\% and 71.89\%, respectively. This choice allows us to retain a relatively large number of sources while ensuring a reliable purity.

\begin{figure}
	\centering
	\includegraphics[width=1.0\linewidth]{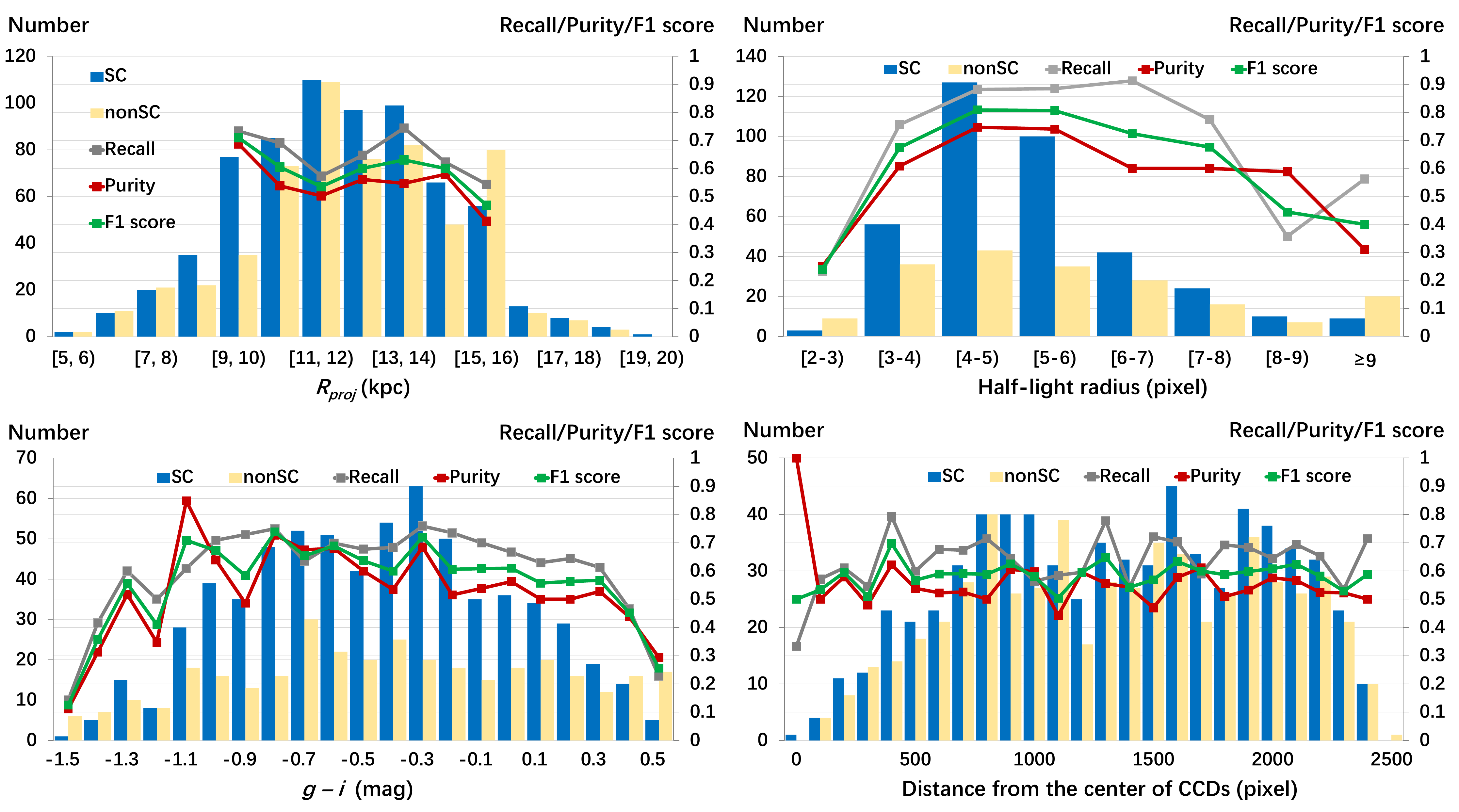}
	\caption{Model performance varies with star cluster projected distance ($R_{proj}$) in kpc (upper-left panel), half-light radius in pixels (upper-right panel), ($g$ - $i$) (lower-left panel), and distance from the center of the CCDs in pixels (lower-right panel).}
	\label{star05_pscd}
\end{figure}

We also evaluate the behavior of our trained model on the test set across various cluster properties, including projected distance, size, color ($g$ - $i$), and distance from the center of CCDs. 
We have divided all test sources into 15 bins based on their $R_{proj}$. In the upper-left panel of Fig.~\ref{star05_pscd}, we plotted the variations in model recall, purity, and F1 score, as well as the number of model predicted clusters. Across these bins, we include both correctly classified true clusters (blue bars) and misclassified entities (yellow bars). From Fig.~\ref{star05_pscd}, it is evident that the recall, purity, and F1 score did not exhibit a declining or increasing trend along with projected distance. Our findings indicate that there does not appear to be a clear relationship between the classification performance and the projected distances from the center of M31. 
From the upper-right panel of Fig.~\ref{star05_pscd}, within the half-light radius range of [$2$\,-\,$3$), the number of star clusters and non-clusters is relatively low, leading to poorer performance across all indicators. This indicates that the model's classification performance in this range is limited. In contrast, when the half-light radius lies within the ranges of [$4$\,-\,$5$) and [$5$\,-\,$6$), the number of star clusters reaches its peak, and the recall, purity, and F1 score are all relatively high, reflecting the model's best performance in these ranges. However, as the half-light radius exceeds 6, the number of star clusters decreases significantly, and all performance indicators decline sharply, resulting in poorer classification performance. 
From the lower-left panel of Fig.~\ref{star05_pscd}, the color within the range of [$-1.1$, $0.3$] corresponds to the peak distribution of star clusters and non-clusters. In this range, the recall, purity, and F1 score also achieve high values, indicating optimal classification performance. At both ends of the ($g$ - $i$) range, classification performance declines, with lower indicator values. This is primarily due to the reduced number of star clusters and non-clusters in these ranges, which limits the model's ability to make accurate classifications.
From the lower-right panel of Fig.~\ref{star05_pscd}, there is no clear direct relationship between pixel distance and classification performance. However, within the pixel distance range of 500 to 2000, the number of star clusters and non-clusters is higher, and the recall, purity, and F1 score remain relatively stable, indicating good classification performance. In contrast, performance at the central and edge regions is poorer, which can be attributed to insufficient sample sizes in these areas.

We have examined the behavior of our star cluster classification model for different subfields within the test region. The results, including accuracy, recall, precision, and F1 score values, as well as the source numbers and projected distances from the center of M31 of the five subfields in the test set, are presented in Table~\ref{5test}. In terms of the number of sources, the R2 sample has the highest number, while the R5 sample has the lowest. Regarding accuracy, all five samples perform exceptionally well. Concerning recall, the R1, R2, R3, and R4 samples exhibit similar performance, while the R5 sample has the lowest recall values. In terms of precision, the R4 sample achieves the highest precision, while the R5 sample has the lowest precision. Moving on to the F1 score, the R4 and R2 samples demonstrate good performance, whereas the R5 sample performs the worst. Overall, the performance of our trained model is similar for the R1 to R4 samples. However, the model demonstrates slightly worse performance in the R5 subfield. This can be attributed to the fact that the R5 subfield is located in the outer disk of M31, where the total number of star clusters is relatively low compared to other subfields. This results in a higher cluster-to-non-cluster ratio and lower recall values, which could negatively impact the performance of the trained model.

\begin{table*}[h]
	\caption{Classification evaluation for the five subfields of the test set.}
	\label{5test}
	\setlength{\tabcolsep}{2.8mm}{
		\renewcommand{\arraystretch}{1.35}
		\begin{tabular}{ccccccc}
			\hline
			Subfield & $R_{proj}$ (kpc) & Source Number & Accuracy & Recall & Precision & F1 score \\
			\hline
			R1 & 11.3845 & 57,634 & 99.71\% & 68.02\% & 54.54\% & 58.65\% \\
			R2 & 10.4647 & 92,159 & 99.69\% & 66.27\% & 55.78\% & 60.57\% \\
			R3 & 10.3948 & 75,103 & 99.74\% & 64.04\% & 51.18\% & 56.89\% \\
			R4 & 12.8544 & 80,952 & 99.78\% & 64.14\% & 65.45\% & 64.79\% \\
			R5 & 16.5834 & 71,480 & 99.82\% & 53.54\% & 38.69\% & 44.92\% \\
			\hline
	\end{tabular}}
\end{table*}

\vspace{-6mm}

\bibliography{sample631}{}

\begin{thebibliography}{}
\expandafter\ifx\csname natexlab\endcsname\relax\def\natexlab#1{#1}\fi
\providecommand{\url}[1]{\href{#1}{#1}}
\providecommand{\dodoi}[1]{doi:~\href{http://doi.org/#1}{\nolinkurl{#1}}}
\providecommand{\doeprint}[1]{\href{http://ascl.net/#1}{\nolinkurl{http://ascl.net/#1}}}
\providecommand{\doarXiv}[1]{\href{https://arxiv.org/abs/#1}{\nolinkurl{https://arxiv.org/abs/#1}}}

\bibitem[{Bialopetravi{\v{c}}ius \&
  Narbutis(2020)}]{bialopetravivcius2020study}
Bialopetravi{\v{c}}ius, J., \& Narbutis, D. 2020, The Astronomical Journal,
  160, 264

\bibitem[{Boulade {et~al.}(2003)Boulade, Charlot, Abbon, Aune, Borgeaud,
  Carton, Carty, Da~Costa, Deschamps, Desforge, {et~al.}}]{boulade2003megacam}
Boulade, O., Charlot, X., Abbon, P., {et~al.} 2003, in Instrument Design and
  Performance for Optical/Infrared Ground-based Telescopes, Vol. 4841, SPIE,
  72--81

\bibitem[{Caldwell {et~al.}(2008)Caldwell, Harding, Morrison, Rose, Schiavon,
  \& Kriessler}]{caldwell2008star}
Caldwell, N., Harding, P., Morrison, H., {et~al.} 2008, The Astronomical
  Journal, 137, 94

\bibitem[{Caldwell {et~al.}(2011)Caldwell, Schiavon, Morrison, Rose, \&
  Harding}]{caldwell2011star}
Caldwell, N., Schiavon, R., Morrison, H., Rose, J.~A., \& Harding, P. 2011, The
  Astronomical Journal, 141, 61

\bibitem[{Cardelli {et~al.}(1989)Cardelli, Clayton, \&
  Mathis}]{cardelli1989relationship}
Cardelli, J.~A., Clayton, G.~C., \& Mathis, J.~S. 1989, Astrophysical Journal,
  Part 1 (ISSN 0004-637X), vol. 345, Oct. 1, 1989, p. 245-256., 345, 245

\bibitem[{Chen {et~al.}(2015)Chen, Liu, Xiang, Yuan, Huang, Huo, Sun, Wang,
  Ren, Zhang, {et~al.}}]{chen2015lamost}
Chen, B.-Q., Liu, X.-W., Xiang, M.-S., {et~al.} 2015, Research in Astronomy and
  Astrophysics, 15, 1392

\bibitem[{Dalal \& Triggs(2005)}]{dalal2005histograms}
Dalal, N., \& Triggs, B. 2005, in 2005 IEEE computer society conference on
  computer vision and pattern recognition (CVPR'05), Vol.~1, Ieee, 886--893

\bibitem[{di~Tullio~Zinn \& Zinn(2014)}]{di2014search}
di~Tullio~Zinn, G., \& Zinn, R. 2014, The Astronomical Journal, 147, 90

\bibitem[{Hannon {et~al.}(2023)Hannon, Whitmore, Lee, Thilker, Deger, Huerta,
  Wei, Mobasher, Klessen, Boquien, {et~al.}}]{hannon2023star}
Hannon, S., Whitmore, B.~C., Lee, J.~C., {et~al.} 2023, Monthly Notices of the
  Royal Astronomical Society, 526, 2991

\bibitem[{He {et~al.}(2016)He, Zhang, Ren, \& Sun}]{he2016deep}
He, K., Zhang, X., Ren, S., \& Sun, J. 2016, in Proceedings of the IEEE
  conference on computer vision and pattern recognition, 770--778

\bibitem[{Hodge {et~al.}(2010)Hodge, Krienke, Bianchi, Massey, \&
  Olsen}]{hodge2010photometric}
Hodge, P., Krienke, O.~K., Bianchi, L., Massey, P., \& Olsen, K. 2010,
  Publications of the Astronomical Society of the Pacific, 122, 745

\bibitem[{Huang {et~al.}(2017)Huang, Liu, Van Der~Maaten, \&
  Weinberger}]{huang2017densely}
Huang, G., Liu, Z., Van Der~Maaten, L., \& Weinberger, K.~Q. 2017, in
  Proceedings of the IEEE conference on computer vision and pattern
  recognition, 4700--4708

\bibitem[{Huxor {et~al.}(2014)Huxor, Mackey, Ferguson, Irwin, Martin, Tanvir,
  Veljanoski, McConnachie, Fishlock, Ibata, {et~al.}}]{huxor2014outer}
Huxor, A., Mackey, A., Ferguson, A., {et~al.} 2014, Monthly Notices of the
  Royal Astronomical Society, 442, 2165

\bibitem[{Johnson {et~al.}(2012)Johnson, Seth, Dalcanton, Caldwell, Fouesneau,
  Gouliermis, Hodge, Larsen, Olsen, San~Roman, {et~al.}}]{johnson2012phat}
Johnson, L.~C., Seth, A.~C., Dalcanton, J.~J., {et~al.} 2012, The Astrophysical
  Journal, 752, 95

\bibitem[{Johnson {et~al.}(2015)Johnson, Seth, Dalcanton, Wallace, Simpson,
  Lintott, Kapadia, Skillman, Caldwell, Fouesneau, {et~al.}}]{johnson2015phat}
---. 2015, The Astrophysical Journal, 802, 127

\bibitem[{Krizhevsky {et~al.}(2012)Krizhevsky, Sutskever, \&
  Hinton}]{krizhevsky2012imagenet}
Krizhevsky, A., Sutskever, I., \& Hinton, G.~E. 2012, Advances in neural
  information processing systems, 25

\bibitem[{Liu {et~al.}(2019)Liu, Zhu, Dai, He, Yao, Tian, Wang, Wu, Zhan, Chen,
  {et~al.}}]{liu2019classification}
Liu, W., Zhu, M., Dai, C., {et~al.} 2019, Monthly Notices of the Royal
  Astronomical Society, 483, 4774

\bibitem[{Lowe(1999)}]{lowe1999object}
Lowe, D.~G. 1999, in Proceedings of the seventh IEEE international conference
  on computer vision, Vol.~2, Ieee, 1150--1157

\bibitem[{McConnachie {et~al.}(2018)McConnachie, Ibata, Martin, Ferguson,
  Collins, Gwyn, Irwin, Lewis, Mackey, Davidge,
  {et~al.}}]{mcconnachie2018large}
McConnachie, A.~W., Ibata, R., Martin, N., {et~al.} 2018, The Astrophysical
  Journal, 868, 55

\bibitem[{Noble(2006)}]{noble2006support}
Noble, W.~S. 2006, Nature biotechnology, 24, 1565

\bibitem[{O'Donnell(1994)}]{o1994rnu}
O'Donnell, J.~E. 1994, Astrophysical Journal, Part 1 (ISSN 0004-637X), vol.
  422, no. 1, p. 158-163, 422, 158

\bibitem[{P{\'e}rez {et~al.}(2021)P{\'e}rez, Messa, Calzetti, Maji, Jung,
  Adamo, \& Sirressi}]{perez2021starcnet}
P{\'e}rez, G., Messa, M., Calzetti, D., {et~al.} 2021, The Astrophysical
  Journal, 907, 100

\bibitem[{Simonyan \& Zisserman(2014)}]{simonyan2014very}
Simonyan, K., \& Zisserman, A. 2014, arXiv preprint arXiv:1409.1556

\bibitem[{Szegedy {et~al.}(2015)Szegedy, Liu, Jia, Sermanet, Reed, Anguelov,
  Erhan, Vanhoucke, \& Rabinovich}]{szegedy2015going}
Szegedy, C., Liu, W., Jia, Y., {et~al.} 2015, in Proceedings of the IEEE
  conference on computer vision and pattern recognition, 1--9

\bibitem[{Viola \& Jones(2004)}]{viola2004robust}
Viola, P., \& Jones, M.~J. 2004, International journal of computer vision, 57,
  137

\bibitem[{Wang {et~al.}(2022)Wang, Chen, Ma, Long, Yuan, Liu, Zhou, Liu, Chen,
  \& He}]{wang2022identification}
Wang, S., Chen, B., Ma, J., {et~al.} 2022, Astronomy \& Astrophysics, 658, A51

\bibitem[{Wang {et~al.}(2023)Wang, Yuan, Chen, Chen, Wu, Niu, Huang, \&
  Liu}]{wang2023searching}
Wang, Y., Yuan, H., Chen, B., {et~al.} 2023, The Astrophysical Journal, 954,
  206

\bibitem[{Wei {et~al.}(2020)Wei, Huerta, Whitmore, Lee, Hannon, Chandar, Dale,
  Larson, Thilker, Ubeda, {et~al.}}]{wei2020deep}
Wei, W., Huerta, E., Whitmore, B.~C., {et~al.} 2020, Monthly Notices of the
  Royal Astronomical Society, 493, 3178

\bibitem[{Zhan(2011)}]{zhan2011consideration}
Zhan, H. 2011, Scientia Sinica Physica, Mechanica \& Astronomica, 41, 1441

\bibitem[{Zhan(2018)}]{zhan2018overview}
---. 2018, 42nd COSPAR Scientific Assembly, 42, E1

\end{thebibliography}
\bibliographystyle{aasjournal}

\end{document}